\documentclass[10pt]{article}

\usepackage{amsmath}
\usepackage{amssymb}

\usepackage{graphicx}

\usepackage[sort&compress]{natbib}
\usepackage{color} 

\usepackage[labelfont=bf,labelsep=period,justification=raggedright]{caption}

\def \cite{\citep}

\date{}

\usepackage{xr}
\usepackage{url}
\graphicspath{{Figures/}}

\newcommand{\figref}[1]{\textbf{Figure~\ref{#1}}}
\newcommand{\tabref}[1]{\textbf{Table~\ref{#1}}}
\newcommand{\supfig}[1]{\textbf{Supplementary Figure~\ref{#1}}}
\newcommand{\suptab}[1]{\textbf{Supplementary Table~\ref{#1}}}
\newcommand{\matmeth}[1]{\textbf{Material and Methods}}
\newcommand{\suptext}[1]{\textbf{Supplementary Text}}

\newcommand{\ignore}[1]{}

\begin{document}

\begin{flushleft}
{\Large
\textbf{Reducing pervasive false positive identical-by-descent segments detected
by large-scale pedigree analysis}
}
\\

Eric Y.\ Durand$^{1,\dagger}$, 
Nicholas Eriksson$^{1}$, 
Cory Y.\ McLean$^{1,\dagger}$
\\

$1$ 23andMe, Inc., Mountain View, CA, USA
\\
$\dagger$ These authors contributed equally to this work.
\\
Corresponding author: Cory Y.\ McLean, cmclean@23andme.com
\end{flushleft}

\clearpage

\section*{Abstract}
Analysis of genomic segments shared identical-by-descent (IBD) between
individuals is fundamental to many genetic applications, from demographic 
inference to estimating the heritability of diseases, but IBD 
detection accuracy in non-simulated data is largely unknown. In principle, 
it can be evaluated using known pedigrees, as IBD segments are by definition 
inherited without recombination down a family tree. We extracted 25,432 
genotyped European individuals containing 2,952 father-mother-child trios from
the 23andMe, Inc. dataset. We then used GERMLINE, a widely used IBD detection 
method, to detect IBD segments within this cohort. Exploiting known familial 
relationships, we identified a false positive rate over 67\% for
2--4 centiMorgan (cM) segments, in sharp contrast with accuracies reported in
simulated data at these sizes. Nearly all false positives arose from the 
allowance of haplotype switch errors when detecting IBD, a 
necessity for retrieving long ($>6$ cM) segments in the presence of 
imperfect phasing. We introduce HaploScore, a novel, computationally efficient
metric that scores IBD segments proportional to the number of switch errors they 
contain. 
Applying HaploScore filtering to the IBD data at a precision of 0.8 produced a
13-fold increase in recall when compared to length-based filtering.
We replicate the false IBD findings and
demonstrate the generalizability of HaploScore to alternative data sources
using an independent cohort of 555 European individuals from the 1000 Genomes
project. HaploScore can
improve the accuracy of segments reported by any IBD detection method,
provided that estimates of the genotyping error rate and switch error rate are
available.

\section*{Introduction}
IBD segments are regions of DNA between two individuals that were inherited from 
a recent shared common ancestor. IBD segments can be detected on
high-density genetic data such as that produced by genome-wide genotyping arrays
or whole genome sequencing.

Detecting the presence and distribution of IBD segments between individuals
is fundamental to many genetic applications~\cite{Browning2012}. Long-range
phasing~\cite{Kong2008}
uses IBD segments to resolve haplotype phasing inaccuracies. IBD segments have
been used to identify disease genes~\cite{Krawitz2010,Gusev2011,Jonsson2012} and estimate
the heritability of traits and common diseases~\cite{Visscher2006,Zuk2012}. 
The lengths and distribution of IBD segments within and across populations have
been used to infer demographic history~\cite{Palamara2012,Gusev2012,Ralph2013} 
and identify regions under natural selection~\cite{Albrechtsen2009,Han2013}.

All methods for IBD detection ultimately try to detect a similarity between 
haplotypes that is statistically unlikely to occur in the absence of IBD sharing.
Hidden Markov models have been used extensively for probabilistic IBD segment
detection~\cite{Purcell2007,Albrechtsen2009,Browning2010,Han2011,Palin2011,Brown2012,Han2013}.
However, these methods scale quadratically with input sample
sizes and are thus not suitable for IBD detection in population-scale datasets
(reviewed in~\cite{Browning2012}).
Non-probabilistic IBD detection methods use a ``hash-and-extend'' methodology
that is conceptually similar to BLAST~\cite{Altschul1990}: identical or 
nearly identical short haplotype match ``seeds'' are detected efficiently and 
the seeds are extended to adjacent sites subject
to heuristic constraints. These non-probabilistic methods have the advantage
that they are able to
scale to much larger datasets than probabilistic methods. Implementations 
include GERMLINE~\cite{Gusev2009,Gusev2012}, fastIBD~\cite{Browning2011}, and
RefinedIBD~\cite{Browning2013}. GERMLINE and RefinedIBD use short windows
of sites as seeds, whereas fastIBD uses small segments of the inferred haplotype
graph as seeds.

These three methods differ in the way that detected candidate segments are
chosen to be kept as true IBD segments: fastIBD uses haplotype frequency,
RefinedIBD uses a combination of segment genetic length and a likelihood ratio
test, and GERMLINE uses segment length.
The probabilistic refinement methods of fastIBD and RefinedIBD require a 
haplotype graph to be generated. Consequently, both fastIBD and RefinedIBD
perform haplotype phasing in addition to IBD detection. 
Haplotype phasing has superlinear computational complexity~\cite{Williams2012}. 
Current computer memory capacity constraints limit the number of individuals who
can be phased together to tens of thousands of individuals. Thus, computing all
pairwise IBD requires splitting the cohort into multiple smaller batches, all of
which must be compared to each other, each time being phased anew. This makes
the detection of all pairwise IBD segments in a cohort of over 100,000
individuals computationally infeasible using these methods. 
Since GERMLINE uses segment length to refine IBD
segments, it does not perform genotype phasing. Consequently, detection of 
all pairwise IBD segments can be performed on large cohorts by phasing each
individual once and then using GERMLINE to detect IBD.

IBD detection accuracy is typically assessed on simulated data, as true IBD
segments can then be known
precisely~\cite{Albrechtsen2009,Browning2007,Gusev2009,Browning2013}.
However, accurate simulation of population demography is
difficult~\cite{Browning2012}, and simulation parameters directly affect the
estimated precision and recall of IBD detection algorithms. With a large number
of father-mother-child trios, IBD detection accuracy can be estimated on 
non-simulated data by examining concordance between reported IBD segments in the
child and his or her parents.

In this work, we analyze the accuracy of IBD segments reported by GERMLINE
since its decoupling of phasing and IBD detection make it feasible for IBD
detection on population-scale datasets. We use a large cohort of trios to assess
IBD segment accuracy on non-simulated data. We perform a detailed examination of
discrepant segments and present a method that substantially improves accuracy
while remaining computationally tractable for population-scale datasets. Finally,
we replicate the findings using an independent cohort of individuals from the 1000
Genomes project.

\section*{Results}
\subsection*{Non-simulated data show substantial inaccuracy in short reported IBD segments}
To analyze IBD detection accuracy on non-simulated data, we examined IBD segments 
detected in a cohort of 25,432 individuals of European ancestry that includes
2,952 distinct father-mother-child trios (the ``23andMe cohort'', \matmeth{XCohortDescription}). By 
focusing specifically on segments reported between a trio child and an 
individual who is not a parent of that child (henceforth called ``child-other''
segments), IBD accuracy can be quantified: by the definition of IBD, if a
child-other segment is true, at least one of the child's parents must also share 
a segment IBD with the individual (henceforth called ``parent-other'' segments) 
that encompasses the child-other segment.

GERMLINE reported a total of 18,125,797 child-other segments in the 23andMe cohort on chromosome 21.
After filtering artifactual IBD segments reported in regions of low site density,
13,307,562 child-other segments were retained for analysis.
Only 14\% of these child-other segments were encompassed by a parent-other segment
(\figref{fig:falseSegs}\textbf{A}, \supfig{fig:parentChoice}\textbf{A}). Another
25\% of child-other segments have a partial parent-other segment in which at
least one segment end is truncated (\figref{fig:falseSegs}\textbf{A},
\supfig{fig:parentChoice}\textbf{B}). Segment ends imply the
presence of opposite homozygote genotypes between the individuals. Opposite
homozygote sites that terminate a parent-other segment exclude the possibility 
of child-other IBD at those sites. To determine whether truncated segment ends
represented false child-other IBD or genotyping error in parent-other regions,
Illumina GenCall scores were examined at the opposite homozygote sites 
truncating 128,656 randomly selected partial parent-other segments. Considering
GenCall scores of $\geq 0.7$ as confident genotype calls~\cite{Fan2003}, over 
95\% of opposite homozygote sites analyzed (122,364/128,656) have confident 
genotype calls in both the parent and other individual. This result indicates
that the vast majority of disagreements between child-other and parent-other
segments represent false positive IBD in the child rather than false negative
IBD in the parent (\figref{fig:falseSegs}\textbf{B}).

The remaining 61\% of child-other segments have no corresponding
parent-other segment (\figref{fig:falseSegs}\textbf{A}, 
\supfig{fig:parentChoice}\textbf{C}).
All segments in this subset were analyzed to determine whether they
represented false positive child-other segments or false negative parent-other
segments by examining the number of parent-other opposite homozygote sites in
the region. Nearly 98\% of these child-other segments have at least one opposite
homozygote site in the parent (\figref{fig:falseSegs}\textbf{C}). Given a 95\%
accuracy rate for parent-other opposite homozygote sites
(\figref{fig:falseSegs}\textbf{B}), the probability that a region
containing $N$ opposite homozygote sites is actually a false negative
parent-other IBD segment was calculated as $(1-0.95)^N$. The expected fraction 
of false negative parent-other segments in this subset is $0.0242$, and thus the
fraction of false positive child-other segments in this subset is $0.9758$. This
likely represents a conservative (i.e.\ low) estimate of false positive
child-other segments for two reasons: the actual genotyping accuracy is much
higher than the stringent confident genotype call threshold indicates, and
segments with no opposite homozygote sites can still be not shared IBD.

\begin{figure}[!ht]
\begin{center}
\includegraphics[width=4in]{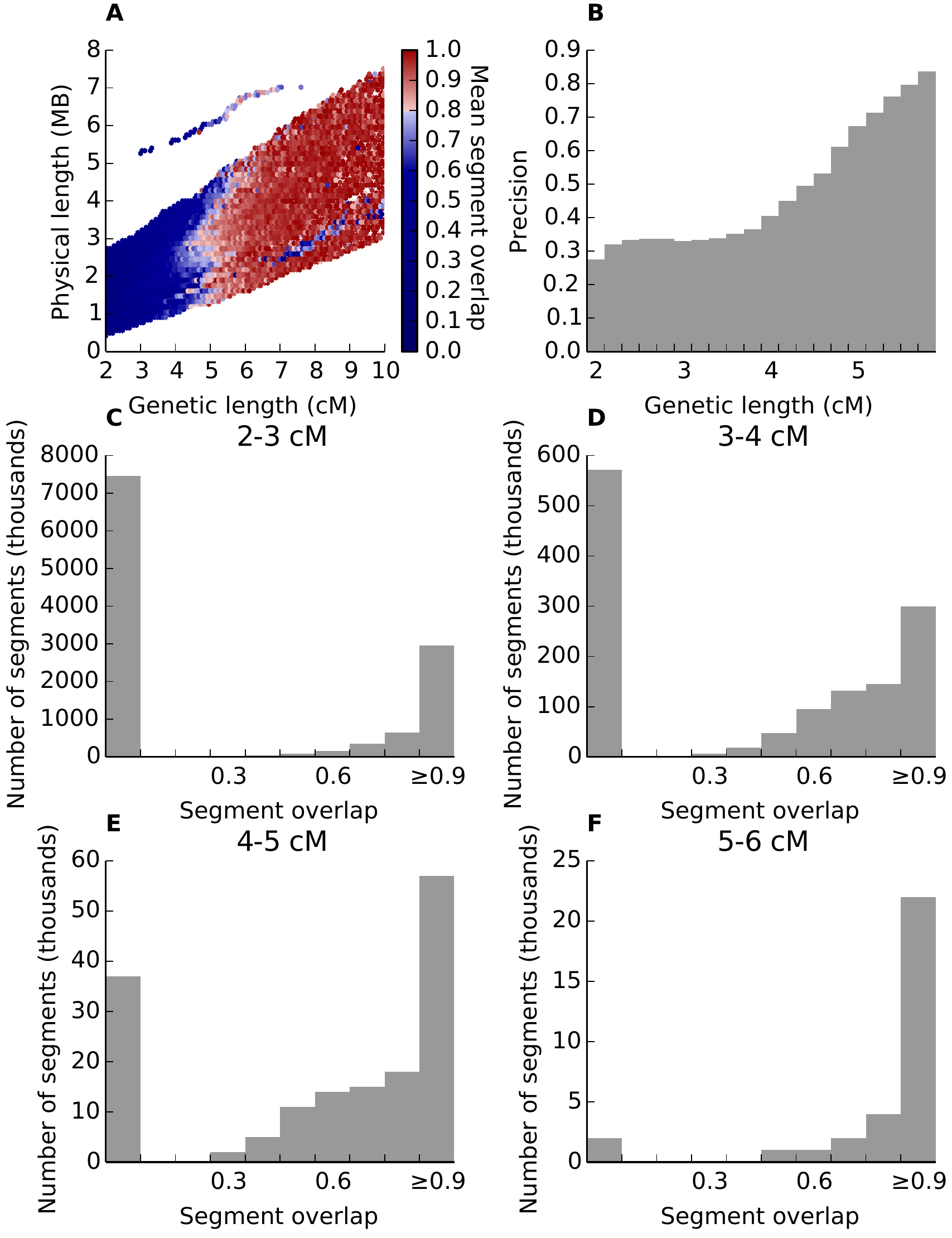}
\end{center}
\caption{
\textbf{Accuracy of child-other IBD segments reported by GERMLINE.}
\textbf{A.} Heat map of the mean fraction of reported child-other IBD segments 
contained in a corresponding parent-other segment, binned by two measures of
segment length. For each child-other segment,
the fraction of the segment also reported as an IBD segment between the parent
and the other individual is calculated. Shown in each bin is the mean of the
segment fractions calculated for all segments in the bin.
\textbf{B.} The fraction of child-other segments that are true IBD as a function of
segment length. True IBD segments are defined as having at least 80\% of their
sites encompassed by a parent-other segment.
\textbf{C--F.} Histograms of child-other segment counts binned by segment
overlap for segments of 2--3 cM (\textbf{C}), 3--4 cM (\textbf{D}), 4--5 cM
(\textbf{E}), and 5--6 cM (\textbf{F}).
Note the scale changes on the y-axes:
though the fraction of true segments of length $<3$ cM is smallest, this range
contains roughly 10-fold more true segments than all other length ranges 
combined.
}
\label{fig:originalPerformance}
\end{figure}

The unexpectedly small number of child-other segments that are fully spanned by
a corresponding parent-other segment motivated an analysis of the relationship
between segment length and segment overlap. Segment overlap between parent and
child was calculated based on the fraction of sites in the child-other segment
(\supfig{fig:parentChoice}). Segments were segregated by genetic and physical
lengths and the average segment overlap of all segments in each bin was
calculated (\figref{fig:originalPerformance}\textbf{A}).
Genetic length is a more reliable indicator of average segment overlap than
physical length and segments longer than 6 cM generally show a
high degree of overlap. However, the average overlap drops rapidly as segment
length is reduced (\figref{fig:originalPerformance}\textbf{A}). 

IBD accuracy was estimated by considering child-other segments with substantial
parent-other segment overlap as true IBD. Because precise determination of IBD
endpoints from genotype data is difficult~\cite{Browning2013}, a threshold of
80\% segment overlap was used to classify a segment as true IBD. Using this
criterion, more than 67\% of all reported segments shorter than 4 cM are false
positive child segments (\figref{fig:originalPerformance}\textbf{B}).
\figref{fig:originalPerformance}\textbf{C}--\textbf{F} show the IBD segment
overlap distributions segregated by genetic length. Most 2--3 cM segments are
erroneous 
(\figref{fig:originalPerformance}\textbf{C}), and only segments longer than 5 cM
have a negligible number of false positives (\figref{fig:originalPerformance}\textbf{F}).
Indeed, when filtering solely by genetic length, all segments shorter than 5 cM
must be discarded in order to achieve a precision value of $0.8$ 
(\supfig{fig:geneticLengthROCs}). 
However, because there are many more short segments, eliminating all segments 
shorter than 5 cM eliminates 99\% of all true IBD segments, a dramatic loss 
in recall (\supfig{fig:geneticLengthROCs}). In the next section, we investigate 
the properties of true IBD segments of all lengths, and contrast them with
erroneous segments. 

\subsection*{Overly permissive diplotype matching causes reported segment inaccuracy}
\label{XWindowAnalysis}
IBD segments are shared between two individual haplotypes. Thus, if the phase 
of each individual genotype was known, IBD detection algorithms could in 
principle analyze each individual haplotype independently. However, for 
individuals without a genotyped pedigree, genotypes have to be phased statistically, 
where switch errors occur at an appreciable frequency (\supfig{fig:switchRates}). 
Examination of only haplotypes in the presence of switch errors is known to 
reduce power to detect IBD, especially for long segments~\cite{Browning2013},
since they are likely to harbor more switch errors than short segments. Thus, GERMLINE (and many other IBD 
detection methods) matches IBD segments between individual diplotypes, trying to
allow for a moderate number of switches between individuals' haplotypes. In practice, 
this is achieved by allowing haplotype match seeds to extend until an opposite
homozygous site is met. There are two potential issues with this approach that
could lead to inconsistent segment reporting between parent and child and are
explored further below.

Detection of child-other segments with a truncated or absent corresponding
parent-other segment could arise from the haplotype matching between the child
and the other individual, but a switch error in the parent causing the 
corresponding haplotype to not match between the parent and the other 
individual. To investigate this potential error source, all 2,952
trios were trio-phased using the laws of Mendelian inheritance and then IBD
detection was performed as before. Trio-phasing ensures that
children and parents are phased essentially perfectly (i.e., up to recombination
events), eliminating haplotype 
discrepancies between parent and child as a source of segment discrepancies.
The number and accuracy of child-other segments using trio-phased data is nearly
identical to that of BEAGLE-phased data, showing that parent-child haplotype 
discrepancies contribute a negligible amount toward discrepant segments
(\supfig{fig:triophased}).

Alternatively, child-other segments with no corresponding parent-other segment 
could be false reported IBD between the child and the other individual due to 
overly permissive diplotype matching. To examine this possibility, each full 
100-site window in all 13,307,562 child-other segments was analyzed (63,542,380 
total windows) to see whether the window satisfied the diplotype match criterion
and the haplotype match criterion between the child and the other individual and
between the parent and the other individual. The analysis was segregated by windows 
contained within corresponding parent-other segments (likely true IBD) and
windows that are not contained within corresponding parent-other segments (false
IBD). The diplotype match criterion is satisfied in the child in 97.6\% of
windows contained within parent-other segments 
(\tabref{tab:windowAnalysis}\textbf{A}) and in 97.5\% of windows not contained 
within parent-other segments (\tabref{tab:windowAnalysis}\textbf{B}). Roughly 
67.7\% of windows contained within both child-other and parent-other segments
satisfy the haplotype match criterion for IBD in the child 
(\tabref{tab:windowAnalysis}\textbf{A}), consistent with true IBD given the 
window size and empirical switch error rate (\supfig{fig:switchRates}). In
contrast, only 44.2\% of windows not contained within a parent-other segment
satisfy the haplotype match criterion for IBD in the child 
(\tabref{tab:windowAnalysis}\textbf{B}), a substantial reduction 
(binomial $P < 10^{-300}$).

The poor precision in short segments is thus due to the allowance of 
diplotype-only matches within the IBD detection algorithm. However, allowing
diplotype-only matches is necessary for detection of long segments due to
imperfect haplotype phasing~\cite{Gusev2012}.
The substantial reduction in windows matching haplotypes in regions of false IBD
suggests a haplotype-based metric that is robust to switch errors could improve 
precision of reported IBD without the loss of recall incurred by haplotype-only
IBD detection mechanisms.

\subsection*{A haplotype-based metric to identify true IBD segments}
IBD is fundamentally a property of haplotypes, not diplotypes. Consequently, 
true IBD should appear consistent with haplotype matches, modulo expected 
genotyping and switch errors. We introduce HaploScore as a measure of haplotype
IBD likelihood: given a genotyping error rate per site $\epsilon$ and a switch 
error rate per site $\sigma$, the HaploScore for a candidate IBD segment $S$ is
\begin{equation*}
\textnormal{HaploScore}(S) = \frac{1}{|S|} \left(\frac{n_g}{\epsilon} + \frac{n_s}{\sigma} \right) \text{,}
\end{equation*}
where $|S|$ is the number of genotyped sites in $S$ and $n_g$ and $n_s$ are the
number of genotyping and switch errors, respectively, that together minimize the
score while reconciling the segment as matching across a single haplotype in
both individuals. Conceptually, HaploScore is a measure of the ratio of observed
and expected genotyping and switch errors. In segments falsely reported as IBD, a 
larger-than-expected number of genotyping and switch errors may be required to 
reconcile the segments as matching across individual haplotypes, and their
HaploScores will be large.

\begin{figure}[!ht]
\begin{center}
\includegraphics[width=3in]{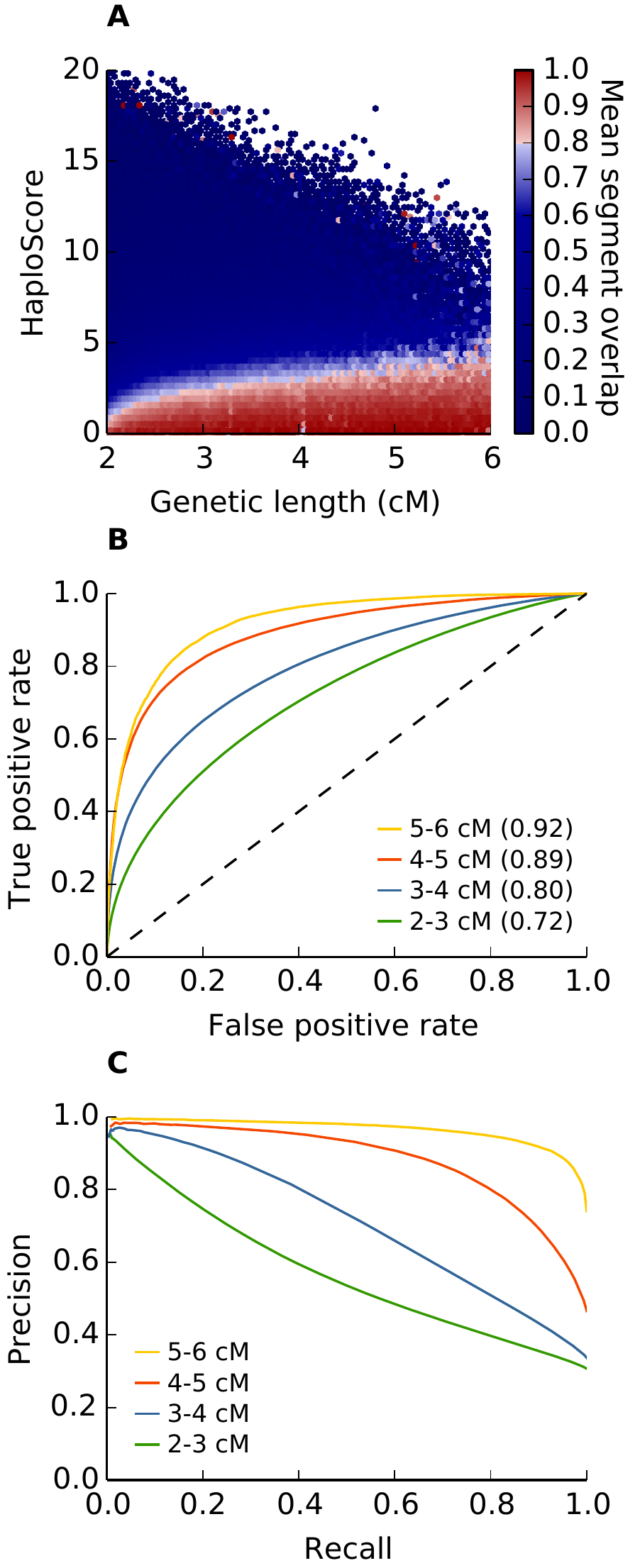}
\end{center}
\caption{
\textbf{Improving detection of true IBD segments using HaploScore.}
\textbf{A.} Heat map of the mean fraction of reported IBD segments found in 
parents, binned by segment genetic length and HaploScore. Calculations are
performed as in~\figref{fig:originalPerformance}\textbf{A}.
\textbf{B.} Receiver operating characteristic for reported IBD segments of
various lengths, discriminating by HaploScore. True IBD is defined as 
in~\figref{fig:originalPerformance}\textbf{B}. The dashed black line indicates 
the no-discrimination line. The area under each curve is parenthesized in its 
legend entry.
\textbf{C.} Precision-recall plot for child-other segments binned by segment
length.
}
\label{fig:haploscorePerformance}
\end{figure}

Genotyping and switch error rates per site were estimated from
the data to be $\epsilon=0.0075$ and $\sigma=0.003$ 
(\matmeth{XParameterEstimation}). Using those parameters, HaploScore was
calculated on all segments shorter than 6 cM. To investigate whether HaploScore 
behaves differently between true and false IBD, 
we plotted a heat map of IBD segment overlap as a function of segment
genetic length and HaploScore values. HaploScore effectively discriminates true 
and false IBD segments at all lengths (\figref{fig:haploscorePerformance}\textbf{A}).
Indeed, the relationship between HaploScore and mean segment overlap is nearly
monotonic, drawing a clear boundary between segments with at least 80\% overlap
and others at all genetic lengths.

In addition, we assessed the power of HaploScore as a binary classifier to
decide if an IBD segment is true. We varied a HaploScore threshold from 0 to 22
(the maximum observed HaploScore value on chromosome 21), and classified segments with a HaploScore
value smaller than the threshold as true IBD. We then computed the true positive
and false positive rates at each HaploScore threshold. HaploScore performed
well as a binary classifier at all genetic lengths, achieving an area under the 
receiver operating characteristic curve (AUC) greater than $0.8$ for segments 
longer than 3 cM (\figref{fig:haploscorePerformance}\textbf{B}). At all levels
of precision, power increased as segment length increased, owing at least in
part to the general positive correlation between segment length and number of
sites in the segment. Importantly, and in sharp contrast with length-based
filtering (\supfig{fig:geneticLengthROCs}), HaploScore-based filtering retains
many segments shorter than 5 cM at a precision of $0.8$
(\figref{fig:haploscorePerformance}\textbf{C}). Recall of HaploScore-based 
filtering at $0.8$ precision is $0.19$, a 13-fold increase compared to
length-based filtering.

\subsection*{Robustness of results to HaploScore parameter variation}
\label{XRobustness}
HaploScore is a function of two parameters: the genotyping error rate $\epsilon$
and the switch error rate $\sigma$. However, only the ratio of the two
parameters affects the behavior of the score. In order to assess the robustness
of HaploScore to varying parameters, a grid search was performed in which 
$\epsilon$ was fixed at $0.0075$, $\sigma$ was varied three orders of magnitude
from $\epsilon/100$ to $10 \epsilon$, and the AUC was computed at each grid 
point (\figref{fig:gridSearches}\textbf{A}). As expected, performance was strongest when
the ratio of the parameters was near its true value. However, the performance 
degradation was modest across the wide range of parameter ratio values examined,
with the AUC dropping by less than 2\% at worst.

\subsection*{Robustness of results to true IBD definition}
\begin{figure}[!ht]
\begin{center}
\includegraphics[width=4in]{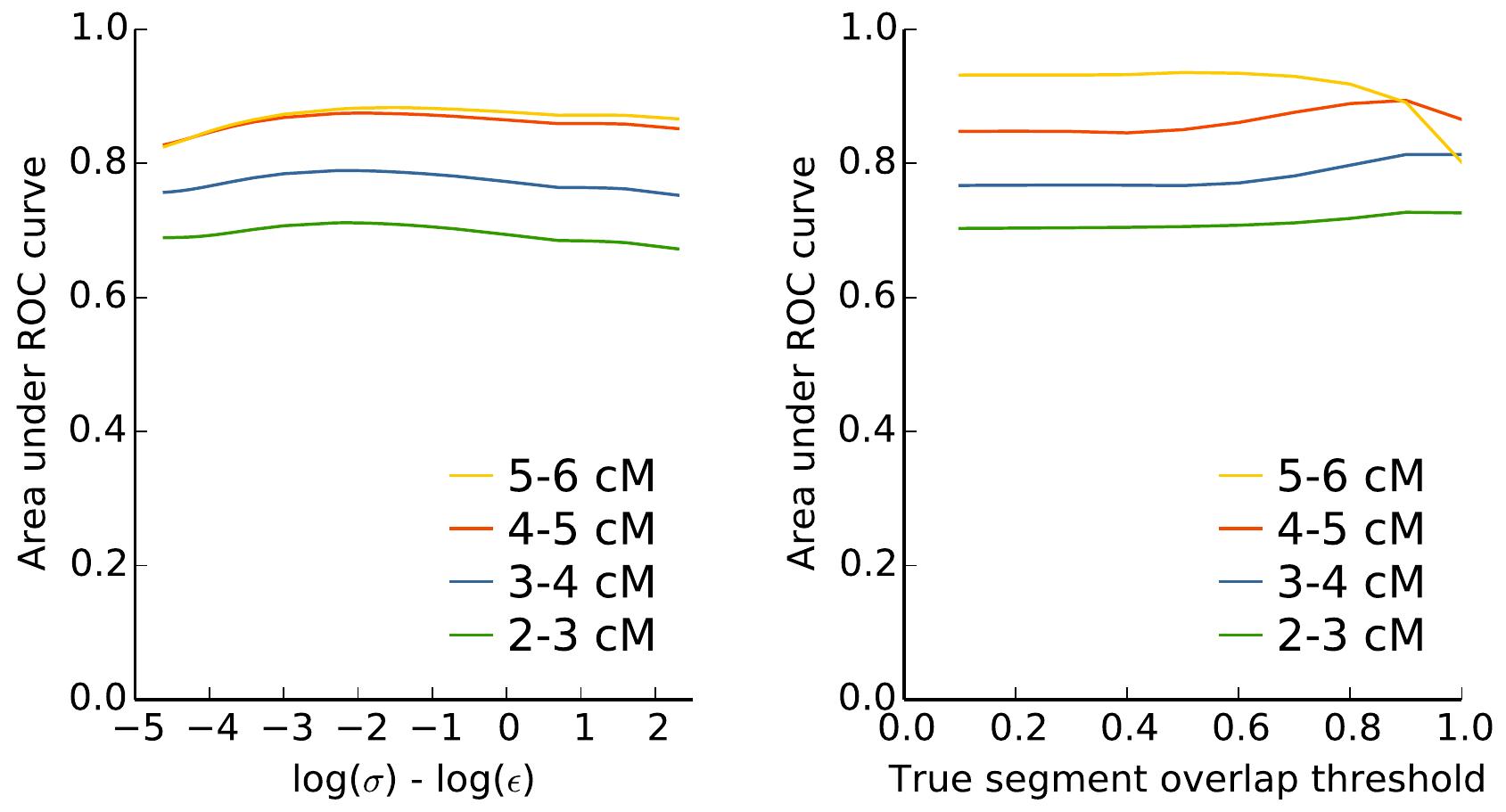}
\end{center}
\caption{
\textbf{HaploScore is robust to a wide range of input parameters.}
\textbf{A.} AUCs for a range of genotyping to switch error rate ratios.
We varied the switch error rate $\sigma$ relative to the genotyping error rate
$\epsilon$. For each value of $\sigma$, we evaluated the resulting AUC
discriminating by HaploScore, where we defined true positive segments as having
a segment overlap of at least $0.80$.
\textbf{B.} AUCs for a range of segment overlap values required
to classify a segment as a true positive.
For each of ten different segment overlap thresholds $(0.1, \ldots, 1.0)$, we
classified true positive segments and calculated the resulting AUC
discriminating by HaploScore.
}
\label{fig:gridSearches}
\end{figure}

In all analyses above, true IBD segments were defined as child-other segments
that have at least 80\% parent-other segment overlap. To assess the robustness
of HaploScore to different true IBD definitions, a grid search was performed in 
which the definition of true IBD was varied from 10\% to 100\% parent-other
segment overlap in increments of 10\%. The AUC was computed at each grid 
point (\figref{fig:gridSearches}\textbf{B}). Performance was generally
stable for all segment lengths and true IBD definitions, with the exception 
of 5--6 cM segments at 100\% overlap, where performance degraded appreciably.
This is likely due at least in part to the inherent bias for longer segments to
have more sites at which premature truncation of detected IBD segments can
arise from genotyping or switch errors.

\subsection*{Robustness of results to genome-wide IBD identification}
To confirm that the results presented are not due to particular genomic features
of chromosome 21, chromosome 10 was analyzed on the full cohort using the same
parameters ($\epsilon=0.0075$, $\sigma=0.003$, 80\% segment overlap defined true
IBD). The results were qualitatively similar to chromosome 21, showing that
the HaploScore methodology is extensible genome-wide
(\supfig{fig:haploChrom10}). In addition, IBD segments were examined on all 
autosomes in the subset of all individuals comprising the 2,952 unrelated trios.
No substantial deviations in performance were observed (not shown).

\subsection*{Filtering spurious reported IBD segments using HaploScore}
HaploScore can be used to filter out spurious segments reported by an IBD
detection algorithm as an efficient post-processing step. The reduced power 
to detect short segments requires more stringent HaploScore 
threshold values for shorter segments to achieve a similar precision
value as for longer segments (\figref{fig:haploscorePerformance}).
Since HaploScore provides a way to rank segments, the trade-off between
precision and recall can be tuned to the needs of the particular downstream 
application.

\begin{figure}[ht]
\begin{center}
\includegraphics[width=4in]{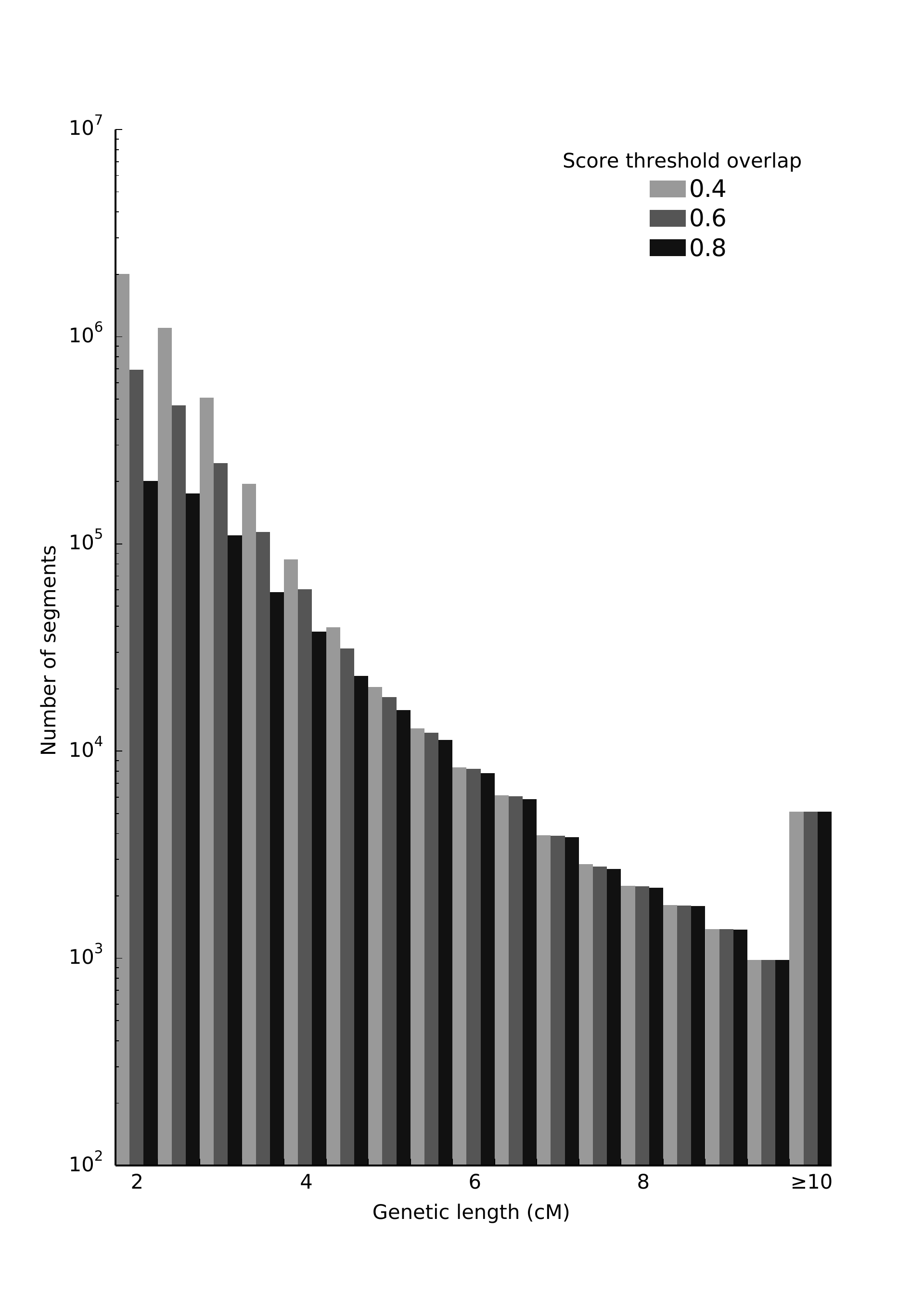}
\end{center}
\caption{
\textbf{Segment detection and HaploScore filtering results.}
Histogram of number of segments reported after filtering at three
different HaploScore thresholds, $t \in \{0.4, 0.6, 0.8\}$. Each threshold $t$
corresponds to a genetic-length-specific array of maximal HaploScores allowed to
retain all segments with mean segment overlap of at least $t$, as described in
the \textbf{HaploScore threshold matrix generation} section of 
\matmeth{XThresholds}. Note that the y-axis is on a log scale.
}
\label{fig:postFilter}
\end{figure}

HaploScore
threshold values to ensure particular average overlap values of resultant 
segments were generated (see~\matmeth{XThresholds}) and three separate filtering results
are shown in~\figref{fig:postFilter}. Notably, more stringent 
filtering parameters have the largest effect on short segments and have nearly
the same effect as lenient filtering parameters for segments over 5 cM
(\figref{fig:postFilter}). This result is intuitive, as the short
reported segments are enriched for false positives
(\figref{fig:originalPerformance}\textbf{C}--\textbf{F}).

\subsection*{Robustness of results to alternate individuals and genotyping platforms}
To assess the robustness of the findings in an alternative population, a cohort
of 555 European individuals including 52 father-mother-child trios genotyped as
part of the 1000 Genomes project~\cite{1000Genomes2012} were analyzed (the 
``1000 Genomes cohort'', \suptab{tab:cohort1000G}). Individuals in the 1000 
Genomes cohort were genotyped on the Illumina HumanOmni2.5-Quad v1-0 B SNP array
and as such provide an independent sample set from which to assess the generalizability
of our results to additional individuals and alternative genotyping platforms.

GERMLINE reported a total of 6,585 child-other segments on chromosome 21 in the
1000 Genomes cohort. After filtering artifactual IBD segments reported in regions
of low site density, 5,770 child-other segments were retained for analysis.
The number of child-other segments detected in the 1000 Genomes cohort is much
smaller than in the 23andMe cohort (5,770 versus
13,307,562 candidate segments) since the 1000 Genomes cohort is much smaller.
However, the rate of candidate segment detection is similar: in the 1000 Genomes
cohort, there are 5,770 segments for $52\times552$ child-other pairs, resulting in
an average of $5770/(52\times552)=0.20$ child-other segments per trio. In the 23andMe
cohort, the corresponding rate is $13307562/(2952\times25429)=0.18$ child-other segments
per trio.

Analyses of child-other segments detected in the 1000 Genomes cohort were performed
analogously to those in the 23andMe cohort. Only
12\% of child-other segments were encompassed by a parent-other segment, 20\% of
child-other segments have a partial parent-other segment in which at least one
segment end is truncated, and the remaining 68\% of child-other segments have no
corresponding parent-other segment (\supfig{fig:falseSegs1000G}\textbf{A}).
Analysis of truncated segments in
the 1000 Genomes cohort also strongly suggests that false child-other IBD 
accounts for most discrepant segments, as 92\% of opposite homozygote sites that
truncate the 1,174 truncated segments have confident genotype calls in both the
parent and other individual (\supfig{fig:falseSegs1000G}\textbf{B}). Finally, in
the 68\% of child-other segments that have no corresponding parent-other segment,
over 99\% contain at least one opposite homozygote site in the parent
(\supfig{fig:falseSegs1000G}\textbf{C}). Taken together, these results show that
the 1000 Genomes cohort is also rife with false positive IBD, and despite the
different genotyping platform used, the error profile in the 1000 Genomes cohort
is qualitatively very similar to that in the 23andMe cohort (\figref{fig:falseSegs}).

Examination of the relationship between segment length and segment overlap in
the 1000 Genomes cohort indicates similar general trends as those discovered in
the 23andMe cohort (compare \supfig{fig:originalPerformance1000G} and
\figref{fig:originalPerformance}), though the smaller number of segments makes
the results more noisy. Comparison of all 44,542 full 100-site windows in the
5,770 child-other segments shows that overly permissive diplotype matching causes
false reported IBD segments: the diplotype match criterion is satisfied in 97.1\%
of windows contained within parent-other segments and in 96.5\% of windows not
contained within parent-other segments, whereas the haplotype match criterion is
satisfied in 68.7\% of windows contained within parent-other segments but in only
51.1\% of windows not contained within parent-other segments
(\suptab{tab:windowAnalysis1000G}), a substantial reduction (binomial 
$P < 10^{-300}$).

Finally, the performance of HaploScore in segregating true and false reported
IBD was analyzed in the 1000 Genomes cohort. The switch error rate was estimated
from the data to be $\sigma=0.003$ and the genotyping error rate was estimated
to be $\epsilon=0.0075$. Similar trends are present in the 1000 Genomes cohort as
are in the 23andMe cohort (compare \supfig{fig:haploscorePerformance1000G} and 
\figref{fig:haploscorePerformance}). The small number of child-other segments analyzed
in the 1000 Genomes cohort causes somewhat noisy results, but the effectiveness
of HaploScore as a discriminator between true and false positive IBD is readily
apparent.

\section*{Discussion}
The usage of IBD segments in genetic analyses will become increasingly common as 
the number of individuals with their genetic composition known increases.
Due to the inherently quadratic nature of IBD detection between all pairs of
individuals in a cohort, non-probabilistic methods are required to keep the
computational burden as low as possible. However, effective filtering methods
are required to ensure reported IBD segments are accurate.

Using the laws of Mendelian inheritance is an effective way to avoid modeling
complex demographic history when evaluating the accuracy of population genetics
methods including IBD detection and local ancestry inference~\cite{Pasaniuc2013}.
By using known familial relationships in a large set of trios, we were able to
analyze the accuracy of IBD segments reported by GERMLINE on non-simulated data.
We found a surprisingly large number of false positive short segments and showed
that these false positives arose due to the diplotype-based IBD detection
mechanism introduced to detect long IBD segments in the presence of
phasing switch errors~\cite{Gusev2012}. We introduced a haplotype-based metric,
HaploScore, that effectively discriminates between true and false reported IBD
segments. We also investigated a likelihood-ratio-based metric, but found it
less effective than HaploScore (\suptext{XLODscore}).

Importantly, HaploScore can be computed efficiently using dynamic 
programming (in O$(|S|)$ time per segment, see~\matmeth{XHaploScoreCalc}). This
suggests a strategy for accurate IBD detection in population-scale datasets:
detect candidate segments using a non-probabilistic IBD detection method with
relatively permissive parameters and then cull true segments using HaploScore
filtering.
In addition, HaploScore can be applied as a post-processing step to existing
genotyping- and sequencing-based IBD segments, provided that an estimate of the
switch error rate and the genotyping error rate are available. 

Achieving optimal HaploScore performance in a different population cohort or
when using an alternative genotyping platform depends on being able to 
accurately estimate the genotyping and switch error rates of the data.
Genotyping error rates can be estimated in any cohort by
methods such as repeat genotyping~\cite{Pompanon2005}. While accurate
determination of switch error rates currently requires trios or orthogonal
analysis methods such as phased sequencing~\cite{Voskoboynik2013}, the
robustness of HaploScore to substantial variations in the parameter ratio
indicates that it should be extensible to non-European populations, genotyping
platforms of different marker density, or even sequencing-based assays. Indeed,
we demonstrated the robustness and generalizability of HaploScore by analyzing an
independent cohort of 555 European individuals from the 1000 Genomes project who
were genotyped on a chip nearly twice as dense as the 23andMe chip. While the
smaller sample size of the 1000 Genomes cohort produced noisier results, all
major findings of the analysis of the 23andMe cohort were replicated in the 1000
Genomes cohort.

Python code implementing HaploScore filtering is freely available
(\url{https://github.com/23andMe/ibd}).

\section*{Material and Methods}
\subsection*{Cohort description}
\label{XCohortDescription}
The 23andMe cohort analyzed comprises 25,432 customers of 23andMe, Inc.,
a personal genetics company, who were genotyped on the Illumina 
HumanOmniExpress+ BeadChip as part of the 23andMe Personal Genome Service. The
chip contains roughly 1,000,000 sites genome-wide~\cite{Hinds2013}. Individuals
were selected for having $>97\%$ European ancestry as described 
previously~\cite{Hinds2013}. The 23andMe cohort includes 2,952 distinct 
father-mother-child trios identified by IBD sharing~\cite{Henn2012}.
Parent-child relationships were defined as having at least 85\% of the genetic 
length of the genome shared on at least one haplotype and no more than 10\% of 
the genetic length of the genome shared on both haplotypes. Parent-parent 
relationships were defined as having at most 20\% of the genetic length of the
genome shared on at least one haplotype.

The 1000 Genomes cohort analyzed comprises 555 individuals from five European
populations who were genotyped on the Illumina HumanOmni2.5-Quad v1-0 B SNP array
as described previously \cite{1000Genomes2012} (samples available at
\url{ftp://ftp-trace.ncbi.nih.gov/1000genomes/ftp/technical/working/20120131_omni_genotypes_and_intensities/Omni25_genotypes_2141_samples.b37.vcf.gz}).
The 1000 Genomes cohort includes 52 distinct father-mother-child trios identified
within the 1000 Genomes project (metadata available at
\url{ftp://ftp-trace.ncbi.nih.gov/1000genomes/ftp/technical/working/20130606_sample_info/20130606_sample_info.txt})
and which we validated independently by IBD sharing (\suptab{tab:cohort1000G}).
All members of the 1000 Genomes cohort were verified to not be present in the
23andMe cohort.

\subsection*{IBD detection}
\label{XIBDDetection}
\subsubsection*{The 23andMe cohort}
Genotypes of all individuals included in the 23andMe cohort were phased using 
BEAGLE~\cite{Browning2007b} version 3.3.1 in batches of 8,000--9,000 individuals
as described previously~\cite{Hinds2013}. In each batch, we excluded sites with
minor allele frequency $<0.001$, Hardy-Weinberg equilibrium $P<10^{-20}$, call
rate $<95\%$, or large allele frequency discrepancies compared to the 1000
Genomes Project reference data. Input haplotypes were restricted to sites
present in the intersection of all batch-filtered sites, and resulted in 12,881
sites on chromosome 21 and 48,372 sites on chromosome 10.

For each of the 2,952 trio children, candidate IBD segments were calculated 
between the child and all 25,429 other individuals who were not the parents of
that child. For each of the 5,904 ($=2 \times 2,952$) trio parents, candidate
IBD segments were calculated between the parent and all 25,430 other individuals
who were not the child of that parent. All candidate IBD segments were 
calculated using the GERMLINE~\cite{Gusev2009} algorithm with the parameters
{\tt -bits 100 -err\_hom 2 -err\_het 0 -w\_extend -min\_m 2 -map <geneticmap>},
corresponding to the empirical genotyping and switch error rates of the data 
(see \textbf{HaploScore parameter estimation} below). The genetic map used was
generated by the Phase II HapMap~\cite{Frazer2007} and lifted over to NCBI Build
GRCh37 coordinates using the UCSC Genome Browser~\cite{Kent2002} liftOver tool 
(available at 
\url{http://hapmap.ncbi.nlm.nih.gov/downloads/recombination/2011-01_phaseII_B37/genetic_map_HapMapII_GRCh37.tar.gz}).
To omit clearly artifactual candidate IBD segments arising from sequence 
assembly gaps and platform effects, candidate segments were filtered by site
density~\cite{Zhuang2012}. Segments with a site density (measured in sites/cM) 
in the lowest 10\% of all 1 cM windows on the chromosome were omitted. All
remaining candidate IBD segments were retained.

\subsubsection*{The 1000 Genomes cohort}
Genotypes of all 555 individuals in the 1000 Genomes cohort were phased using
BEAGLE~\cite{Browning2007b} version 3.3.1 in a single batch. Windows of 3,000
sites that overlapped by 100 sites were stitched together as described
previously~\cite{Hinds2013}. Sites that were not polymorphic in the 555
individuals examined, had a 1000-Genomes-reported Hardy-Weinberg equilibrium
$P<10^{-20}$, or a call rate within the 555 individuals examined $<95\%$ were
excluded, resulting in 23,142 sites on chromosome 21. GenCall genotype scores
were set to 0 for all sites not called in each individual.

Candidate IBD segments were identified and filtered identically to those found
in the 23andMe cohort described above.

\subsection*{HaploScore description and computational complexity}
\label{XHaploScoreCalc}
HaploScore provides a metric by which to rank the likelihood that a stretch of
DNA is inherited IBD between two individuals or not. Let $\epsilon$ and $\sigma$
denote the probability of a genotyping error and a switch error at any 
given site, respectively. The HaploScore for a candidate IBD segment $S$ is
\begin{equation}
\textnormal{HaploScore}(S) = \frac{1}{|S|} \left(\frac{n_g}{\epsilon} + \frac{n_s}{\sigma} \right)
\end{equation}
where $|S|$ is the number of genotyped sites in $S$ and $n_g$ and $n_s$ are the 
number of genotyping and switch errors, respectively, that together minimize the
score while reconciling the segment as matching across a single haplotype in
both individuals.

Finding the HaploScore (i.e.\ the optimal values of $n_g$ and $n_s$ subject to
the constraints) can be viewed as finding the minimum-cost path through the 
directed acyclic graph (DAG) described below (\supfig{fig:haploGraph}).

Let $G$ be a DAG with a single source node and a single sink node. Between the
source and the sink, the graph has $|S|$ levels, one per genotyped site in 
segment $S$. Each of these $|S|$ levels has four nodes, one for each possible
haplotype configuration. Each node in level $l$ has four outgoing directed 
edges, one to each node in level $l+1$. Below, we use the same notation for
nodes and their weights.

At any level $l$, let $h_l^{(i,j)}$, $i,j \in \{1, 2\}$ denote the four possible 
haplotype configurations of an IBD match. The nodes are weighted as follows:

\begin{equation}
h_l^{(i,j)} = \begin{cases}
  0 & \text{if haplotype } i \text{ in first individual matches haplotype } j \text{ in second individual at position } l \text{,}\\
  1/\epsilon & \text{otherwise.}
\end{cases}
\end{equation}

Let $e_l^{(i,j),(u,v)}$ denote the weight of the edge between nodes $h_l^{(i,j)}$
and $h_{l+1}^{(u,v)}$. Edges are weighted as follows:

\begin{equation}
e_l^{(i,j),(u,v)} = \begin{cases}
  0          & \text{if } i=u \text{ and } j=v \text{,}\\
  1/\sigma   & \text{if } i=u \text{ and } j \neq v \text{,}\\
  1/\sigma   & \text{if } i \neq u \text{ and } j=v \text{,}\\
  2/\sigma   & \text{if } i \neq u \text{ and } j \neq v \text{.}
\end{cases}
\end{equation}

The weights of the four edges from the source node to the nodes in the first
level, as well as the weights from the nodes in level $|S|$ to the sink node,
are set to 0. The cost of a path in $G$ is defined as the sum of the weights of
the edges and nodes it traverses.

HaploScore$(S)$ is equal to the smallest of all path costs from the source to 
the sink. It can be efficiently computed using dynamic programming by noting 
that the smallest cost from the source to level $l+1$ in the graph can easily be
inferred from the smallest cost from the source to level $l$. Let $C_l^{(i,j)}$
denote the smallest cost from the source to haplotype configuration $(i,j)$ at
level $l$. Then,

\begin{equation}
C_{l+1}^{(u,v)} = \min_{i,j,u,v} \left(C_l^{(i,j)} + e_l^{(i,j),(u,v)} + h_{l+1}^{(u,v)} \right) \text{.}
\end{equation}

The minimum cost to reach level $l$, $C_l^*$, is then

\begin{equation}
C_l^* = \min_{i,j} C_l^{(i,j)} \text{.}
\end{equation}

The above equations clearly show that computing HaploScore$(S)$ involves 16
comparisons at each genotyped site in $S$. Thus, the complexity of computing 
HaploScore$(S)$ is at most $16|S|$. Performance can be further improved when
filtering by HaploScore by terminating computation as soon as a segment's 
HaploScore becomes too high to satisfy the maximum value threshold.

\subsection*{HaploScore parameter estimation}
\label{XParameterEstimation}
HaploScore uses two parameters, the genotyping error rate per site $\epsilon$ 
and the switch error rate per site $\sigma$. Analyses of genotyping chip
accuracy~\cite{Paynter2006} and internal comparisons between genotype and
whole-genome sequencing data verify that the genotyping error rate is $<1\%$
(not shown). To estimate the empirical switch error rate per site, all 2,952
trios in the 23andMe cohort were trio-phased using the laws of Mendelian
inheritance and the results
for all children were compared to their BEAGLE-phased haplotypes, assuming 
that the trio-phased haplotypes represented the true phase. The average per-site
switch error rate ranged from $0.0019$ (on chromosome 6) to $0.0043$ (on 
chromosome 19) but deviated only modestly from a constant rate on each
chromosome (\supfig{fig:switchRates}).

The switch error rate calculation process described above was performed 
independently on the 1000 Genomes cohort. A total of 3,629 switch errors were
detected in the 52 trio children over 23,142 sites. This corresponds to an 
individual switch error rate per site of $3629/(52\times23142) = 0.003$.

\subsection*{HaploScore threshold matrix generation}
\label{XThresholds}
A matrix of HaploScore thresholds was generated in the following manner:  all 
segments were binned by genetic length in $0.1$ cM increments from $2$ cM to 
$10$ cM. In each length bin, segments were segregated by their segment overlap
into $100$ equally-sized overlap bins. The score threshold in each overlap bin
was initially set to be the average HaploScore of all segments within the bin. 
To ensure monotonicity, the score threshold was then taken to be the maximum of
the scores in all bins of equal or higher overlap at that segment length.
A file containing the maximum HaploScore value thresholds calculated for all
genetic lengths and mean overlap values in the 23andMe cohort is available
(\textbf{Supplementary File 1}).

\section*{Acknowledgments}
We thank the customers of 23andMe who contributed the genetic data that made
this research possible and are grateful to the employees of 23andMe for creating
and supporting the resources necessary for this research. We also thank members
of the 23andMe research team for insightful comments.
This work was supported by the National Human Genome Research Institute of
the National Institutes of Health (grant number R44HG006981).
%

\bibliography{ibd}

\clearpage
\begin{samepage}
\section*{Figures}
\begin{figure}[!ht]
\begin{center}
\includegraphics[width=4in]{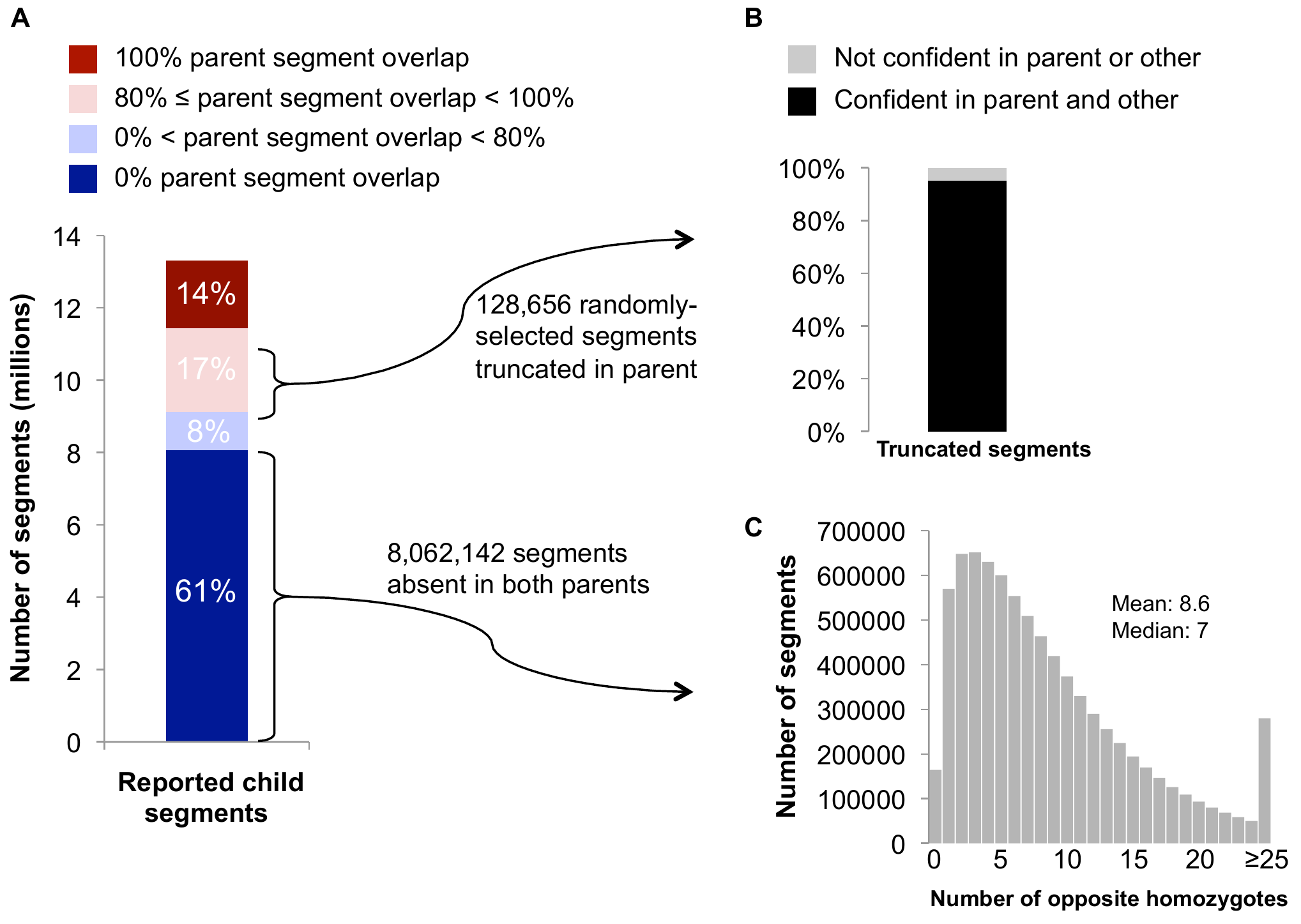}
\end{center}
\caption{
\textbf{Analysis of child-other segments in parents.}
\textbf{A.} The majority of child-other segments are not detected in either parent.
Parent segment overlap is calculated as the percentage of sites in the 
child-other segment that are included in the parent-other segment.
\textbf{B.} Truncation points for parent-other segments are nearly always
confidently-genotyped opposite homozygote sites, consistent with false positive 
IBD in the child. The opposite homozygote site causing truncation of the 
parent-other segment was examined in a randomly-selected subset of all 3,371,616
segments with partial parent overlap.
\textbf{C.} Child-other segments with no corresponding parent-other segments contain many 
parent-other opposite homozygotes in the region, also consistent with false 
positive IBD in the child. For each of these child-other segments, 
the number of opposite homozygote sites present between the parent and the other 
individual at that segment location is calculated separately for each parent, 
and the smaller is chosen as the number of opposite homozygotes in the region.
}
\label{fig:falseSegs}
\end{figure}
\end{samepage}

\clearpage
\begin{samepage}
\section*{Tables}
\begin{table}[!ht]
\caption{
\textbf{Haplotype and diplotype window matches in child-other segments.}
\textbf{A.} Counts of window types in windows contained within a corresponding
parent-other segment.
\textbf{B.} Counts of window types in windows that are not contained within a
corresponding parent-other segment.
}
{\large \textbf{A}} \\
\begin{tabular}{lrrrr}
                  & \textbf{Child Diplo} & \textbf{Child Haplo} & \textbf{Child Both} & \textit{\textbf{Total}} \\
\hline                  
\textbf{Par None}    &         ~~~~~~~~0 &          ~~~~~~0  &      ~~~~~~~~~0  &     \textit{~~~~~~~~~0} \\
\textbf{Par Diplo}   &         6,283,300 &          ~56,393  &      ~1,045,425  &     \textit{~7,385,118} \\
\textbf{Par Haplo}   &         ~~~57,353 &          243,447  &      ~~ 236,157  &     \textit{~~~536,957} \\
\textbf{Par Both}    &         1,098,359 &          243,490  &      13,733,586  &     \textit{15,075,435} \\
\hline
\textit{\textbf{Total}} &   \textit{7,439,012} &  \textit{543,330}  &  \textit{15,015,168}  &     \textit{22,997,510} \\
\end{tabular} \\
{\large \textbf{B}} \\
\begin{tabular}{lrrrr}
                  & \textbf{Child Diplo} & \textbf{Child Haplo} & \textbf{Child Both} & \textit{\textbf{Total}} \\
\hline
\textbf{Par None}    &        14,055,602 &         ~483,921  &      ~5,853,193  &     \textit{20,392,716} \\
\textbf{Par Diplo}   &        ~7,574,059 &         ~~77,905  &      ~2,068,399  &     \textit{~9,720,363} \\
\textbf{Par Haplo}   &        ~~~~82,378 &         ~243,698  &      ~~~372,885  &     \textit{~~~698,961} \\
\textbf{Par Both}    &        ~~~931,599 &         ~222,127  &      ~8,579,104  &     \textit{~9,732,830} \\
\hline
\textit{\textbf{Total}} &  \textit{22,643,638} &  \textit{1,027,651}  & \textit{16,873,581}  &  \textit{40,544,870} \\
\end{tabular}
\begin{flushleft}
Par, parent; Diplo, diplotype match only; Haplo, haplotype match only.
\end{flushleft}
\label{tab:windowAnalysis}
\end{table}
\end{samepage}
\clearpage

\renewcommand\thefigure{S\arabic{figure}}
\setcounter{figure}{0}
\renewcommand\thetable{S\arabic{table}}
\setcounter{table}{0}

\section*{Supplementary Note}
\subsection*{Logarithm of Odds (LOD) segment scoring}
\label{XLODscore}
In this section, we describe an alternative scoring for potential IBD segments
that is similar in spirit to the LOD score used in
RefinedIBD~\cite{Browning2013}. Specifically, for a given segment $S$ shared
between two individuals $i_1$ and $i_2$, we compute its LODscore as follows:

\begin{equation}
\textnormal{LODscore}(S) = log \left(\frac{Pr \left(G_{obs1}^{(S)}, G_{obs2}^{(S)} | \text{IBD} \right)}{Pr \left(G_{obs1}^{(S)}, G_{obs2}^{(S)} | \text{no IBD} \right)} \right) \text{,}
\end{equation}

\noindent
where $G_{obs1}^{(S)}$ (resp.\ $G_{obs2}^{(s)}$) is the observed genotype of
individual $i_1$ (resp.\ $i_2$) over segment $S$, and
$Pr \left(G_{obs1}^{(S)}, G_{obs2}^{(S)} | \text{IBD} \right)$
(resp.\ $Pr \left(G_{obs1}^{(S)}, G_{obs2}^{(S)} | \text{no IBD} \right)$)
is the pseudo-likelihood of observing $G_{obs1}^{(S)}$ and $G_{obs2}^{(S)}$
conditioned on individuals $i_1$ and $i_2$ being IBD over segment $S$ (resp.\
not being IBD).

The pseudo-likelihood is computed as follows:

\begin{eqnarray*}
\lefteqn{Pr\left(G_{obs1}^{(s)}, G_{obs2}^{(s)} | \text{IBD}, \epsilon \right) = } \\
&& \prod_{i=1}^{\#S} \sum_{G_{true1}} \sum_{G_{true2}} Pr\left(G_{true1}^{(i)},G_{true2}^{(i)} | \text{IBD}\right) Pr\left(G_{true1}^{(i)} | G_{obs1}^{(i)}, \epsilon \right) Pr\left(G_{true2}^{(i)} | G_{obs2}^{(i)}, \epsilon \right) \text{,}
\end{eqnarray*}

\noindent
where $\epsilon$ is the genotyping error rate, $\#S$ is the number of markers in
the IBD segment, and $G_{true\textnormal{j}}^{(i)}$ is the true genotype of
individual j at position $i$. The probability of genotypes
$(G_{true1}^{(i)}, G_{true2}^{(i)})$ as a function of the IBD state (0, 1 or 2
alleles shared IBD at position $i$) is given in~\suptab{tab:ibdgenoprobs}. We note
that~\suptab{tab:ibdgenoprobs} was derived elsewhere~\cite{Albrechtsen2009}. The
probability of observing a genotype given the true genotype and the genotyping 
error rate is given in~\suptab{tab:observedgenoprobs}. Two genotypes are
considered IBD if they either share one or two alleles IBD (IBD1 and IBD2
in~\suptab{tab:observedgenoprobs}), and we give equal prior probabilities to the
two configurations.

\begin{figure}[!ht]
\begin{center}
\includegraphics[width=4in]{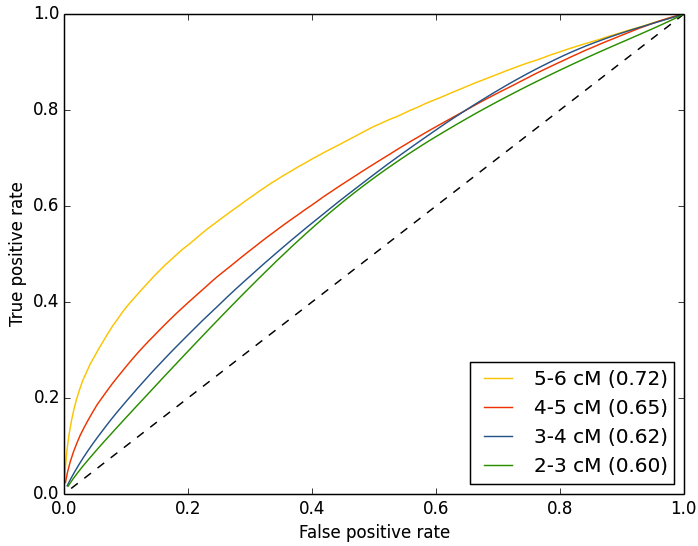}
\end{center}
\caption{
\textbf{Receiver operating characteristic for reported IBD segments of various 
lengths, discriminating by LODscore.}
True positive IBD segments are defined as having at least 80\% of their sites
encompassed by a parent-other segment. The area under each curve is 
parenthesized in its legend entry.
}
\label{fig:lodscore}
\end{figure}

We assessed the performance of LODscore by computing its AUC for various segment
sizes. We note that even though the LODscore has power to filter out false IBD 
segments, its AUC is generally lower than the HaploScore detailed in the main 
text (\supfig{fig:lodscore}). Reasons for the lower power of LODscore may arise 
in part from two issues:  1) LODscore assumes each site is independent and thus
ignores correlation between adjacent markers, and 2) LODscore ignores available
phase information. Both issues could be alleviated by explicitly incorporating
linkage disequilibrium between adjacent sites and switch errors into the model.
However, because of the strong performance of HaploScore, we did not explore
these research avenues further.

\clearpage
\begin{samepage}
\section*{Supplementary Figures}
\begin{figure}[!ht]
\begin{center}
\includegraphics[width=4in]{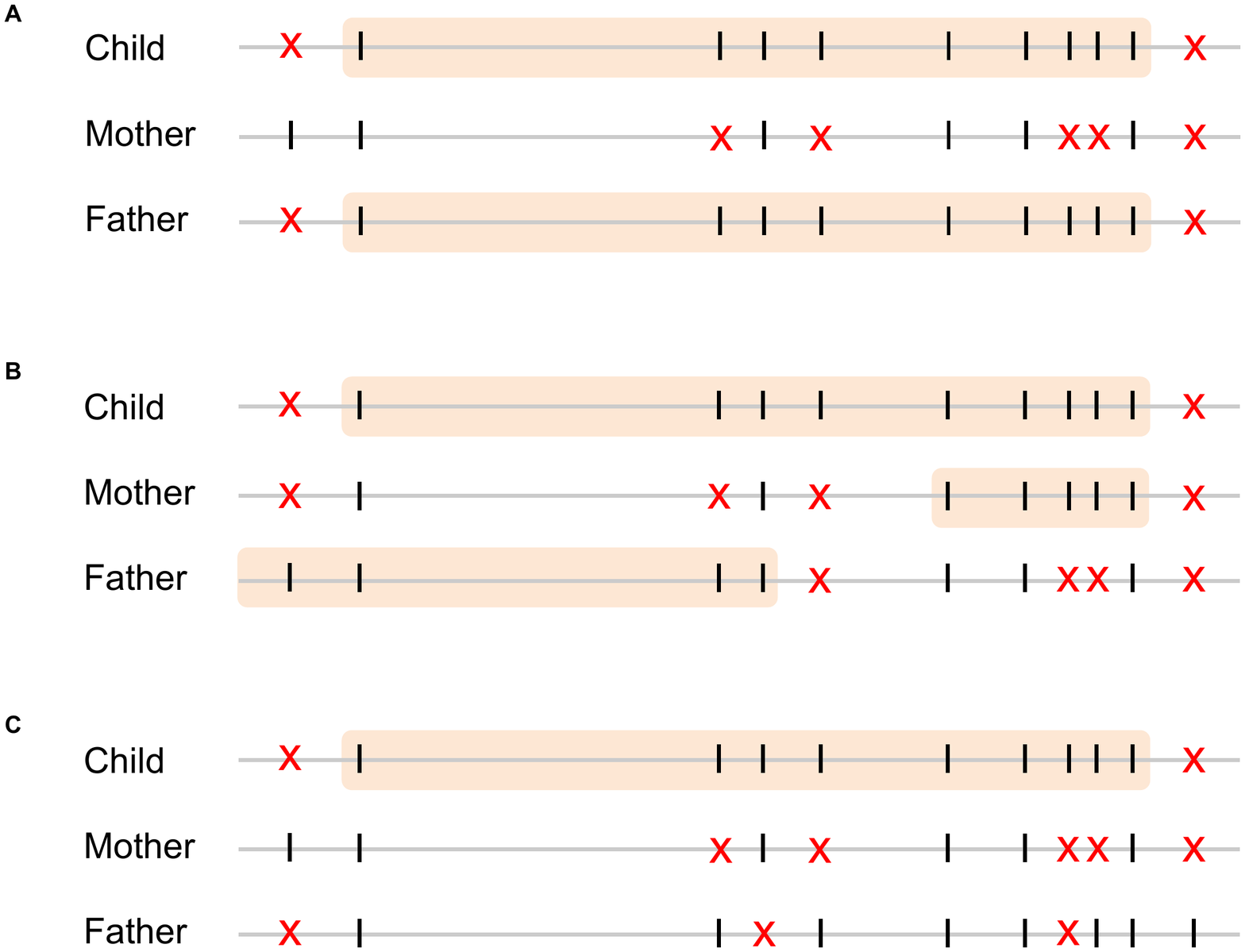}
\end{center}
\caption{
\textbf{Choosing the parent through which child-other IBD segments have been
transmitted.}
The genome is represented as a horizontal gray line. Assayed sites compatible
with IBD between the listed individual and a hypothetical other individual (not
pictured) are indicated as vertical black lines. Assayed sites incompatible with
IBD (e.g., opposite homozygote sites) are indicated as red crosses. Orange boxes
indicate reported IBD segments between the listed individual and the
hypothetical other individual (not pictured).
\textbf{A.} The unambiguous case in which one parent has a corresponding IBD 
segment and the other parent does not. Here, the father would be selected as the
parent for analysis.
\textbf{B.} The case where each parent has an IBD segment that partially overlaps 
the child segment. The parent selected for analysis is determined by the 
fraction of sites shared IBD. In this case, despite the longer physical length
of the father's segment, the mother would be selected since her segment overlap
(5 of 9 sites) is larger than the father's (3 of 9 sites).
\textbf{C.} The case where neither parent has a reported IBD segment. The father 
would be selected as the parent for analysis, since his genotype contains fewer
opposite homozygote sites in the child IBD region.
}
\label{fig:parentChoice}
\end{figure}
\end{samepage}

\begin{figure}[!ht]
\begin{center}
\includegraphics[width=4in]{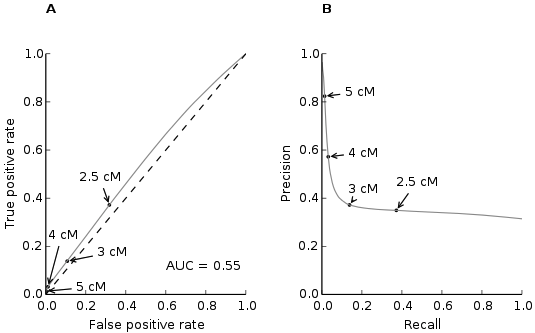}
\end{center}
\caption{
\textbf{Receiver operating characteristics of child-other IBD segments
discriminating by genetic length.}
\textbf{A.} True positive rate vs.\ false positive rate when discriminating by
minimum genetic length.
\textbf{B.} Precision vs.\ recall when discriminating by minimum genetic length.
Values for four particular minimum genetic length criteria are marked on each
plot.
}
\label{fig:geneticLengthROCs}
\end{figure}

\begin{figure}[!ht]
\begin{center}
\includegraphics[width=4in]{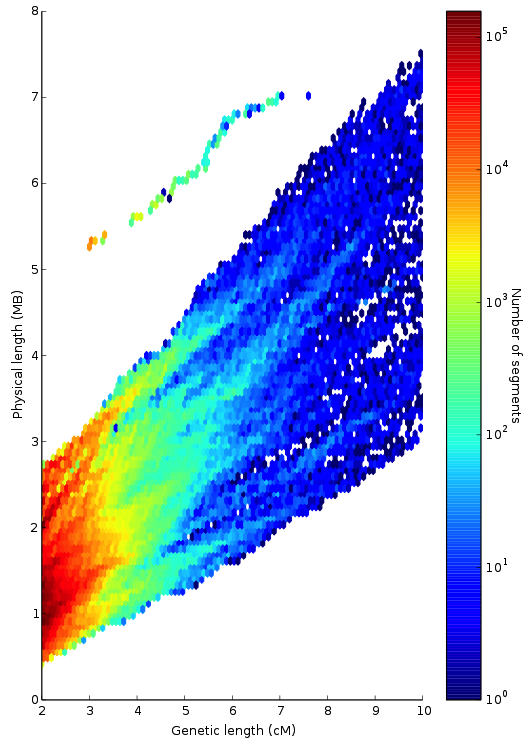}
\end{center}
\caption{
\textbf{Length distribution of child-other IBD segments.}
Heat map shows the number of segments in each bin segregating by the genetic and
physical lengths of the segments. Axes identical to those 
in~\figref{fig:originalPerformance}\textbf{A}.
}
\label{fig:lengthHeatmap}
\end{figure}

\begin{figure}[!ht]
\begin{center}
\includegraphics[width=4in]{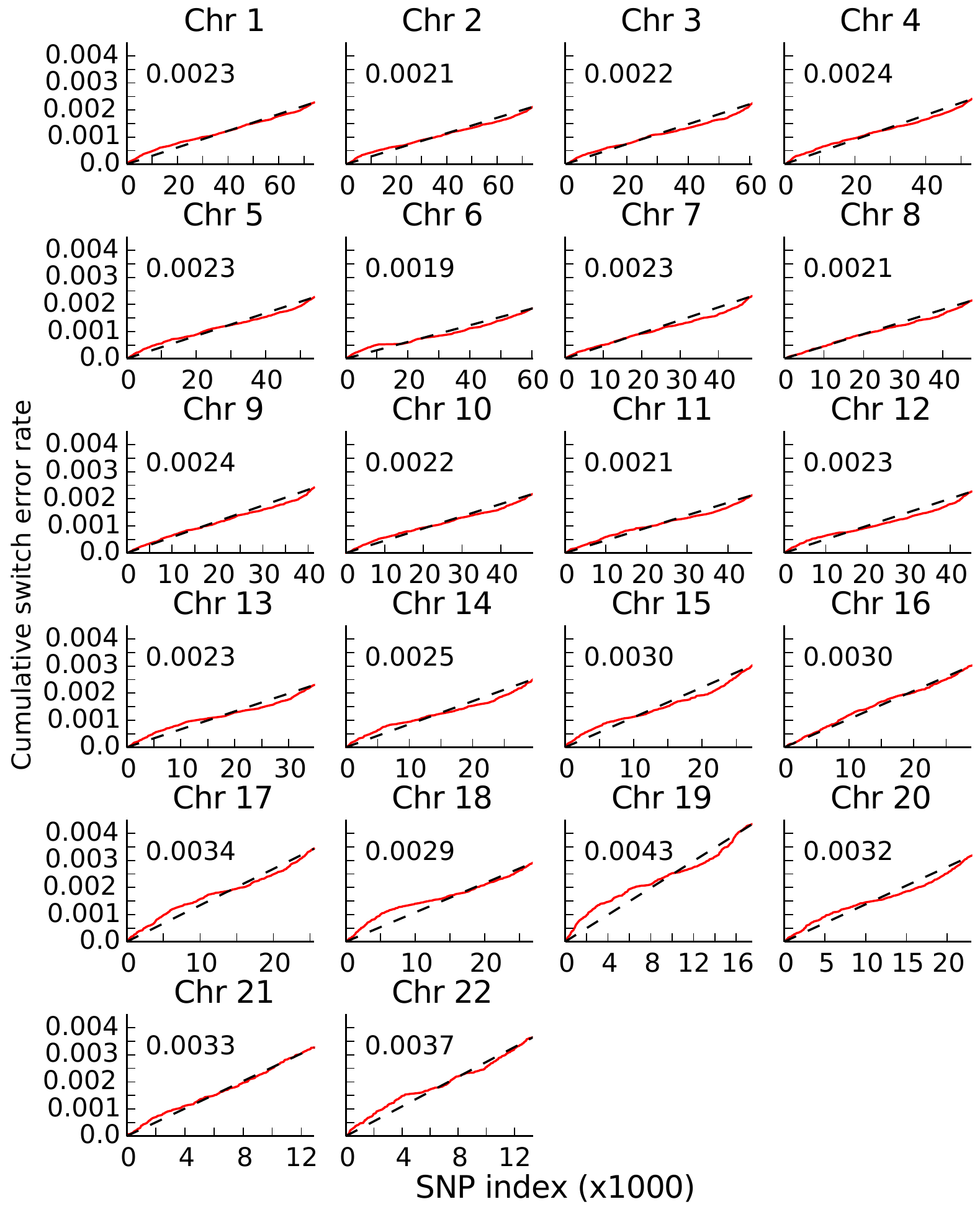}
\end{center}
\caption{
\textbf{Switch errors in BEAGLE-phased data occur at a nearly-constant rate across
chromosomes.}
Switch error positions were
detected in 2,952 trio children by comparing BEAGLE-phased haplotypes with 
trio-phased haplotypes and assuming trio-phased data was truth. The average
individual switch error rate was calculated at each site by dividing the total
number of switch errors at that site by 2,952. Red lines plot the cumulative 
switch error rate scaled by the number of sites on the chromosome, to 
facilitate inter-chromosomal comparison. Numbers in the top left of each graph
indicate the average per-site switch error rate for the chromosome. Black dashed
lines indicate the expected individual cumulative switch errors per site 
assuming a constant switch error rate at each site on the chromosome.
}
\label{fig:switchRates}
\end{figure}

\begin{figure}[!ht]
\begin{center}
\includegraphics[width=4in]{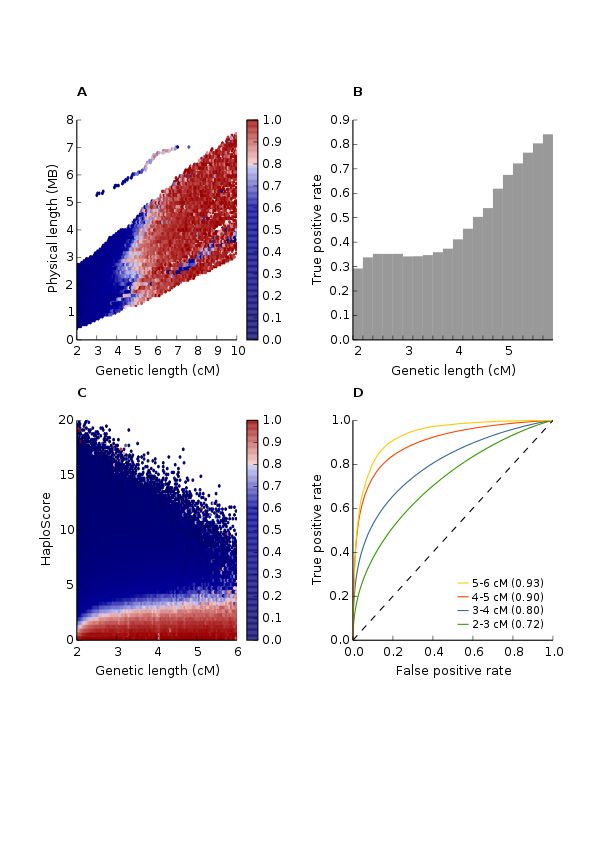}
\end{center}
\caption{
\textbf{IBD segment overlap and HaploScore performance on chromosome 21 using
trio-phased trios.}
\textbf{A.} Heat map of the mean fraction of reported IBD segments found in 
parents, binned by two measures of segment length.
\textbf{B.} The fraction of child-other segments that are true IBD as a function
of segment length. True IBD segments are defined as having at least 80\% of
their sites encompassed by a parent-other segment.
\textbf{C.} Heat map of the mean fraction of reported IBD segments found in
parents, binned by segment genetic length and HaploScore.
\textbf{D.} Receiver operating characteristic for reported IBD segments of
various lengths, discriminating by HaploScore. The four panels are analogous 
to~\figref{fig:originalPerformance}\textbf{A},\textbf{B} 
and~\figref{fig:haploscorePerformance}\textbf{A},\textbf{B}, respectively, 
using trio-phased data for all 2,952 trios. The similarity of this figure and 
the main text figure panels indicates that haplotype phasing errors do not 
contribute substantially to the estimates of IBD accuracy.
}
\label{fig:triophased}
\end{figure}

\begin{figure}[!ht]
\begin{center}
\includegraphics[width=4in]{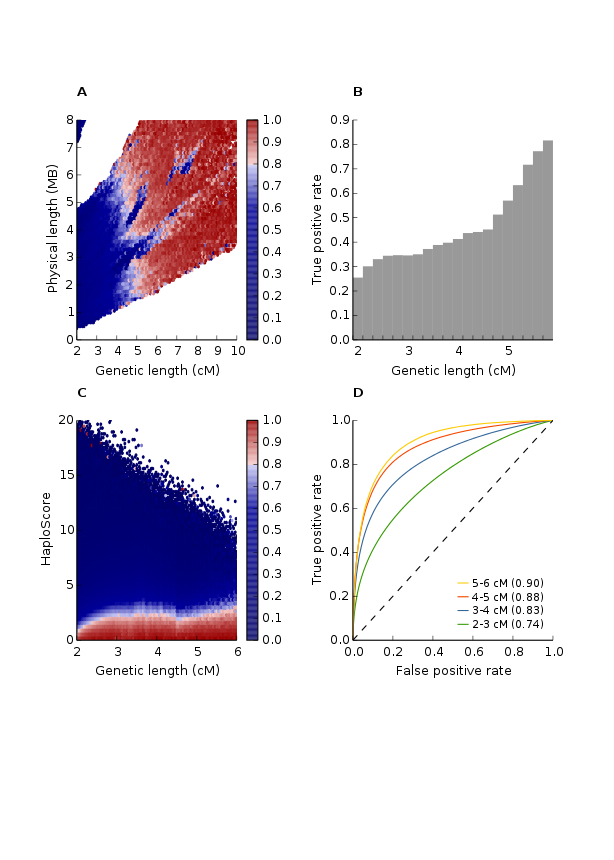}
\end{center}
\caption{
\textbf{IBD segment overlap and HaploScore performance on chromosome 10.}
\textbf{A.} Heat map of the mean fraction of reported IBD segments found in 
parents, binned by two measures of segment length.
\textbf{B.} The fraction of child-other segments that are true IBD as a function
of segment length. True IBD segments are defined as having at least 80\% of
their sites encompassed by a parent-other segment.
\textbf{C.} Heat map of the mean fraction of reported IBD segments found in
parents, binned by segment genetic length and HaploScore.
\textbf{D.} Receiver operating characteristic for reported IBD segments of
various lengths, discriminating by HaploScore. The four panels are analogous 
to~\figref{fig:originalPerformance}\textbf{A},\textbf{B} 
and~\figref{fig:haploscorePerformance}\textbf{A},\textbf{B}, respectively, 
calculated on chromosome 10 here.
}
\label{fig:haploChrom10}
\end{figure}

\begin{figure}[!ht]
\begin{center}
\includegraphics[width=4in]{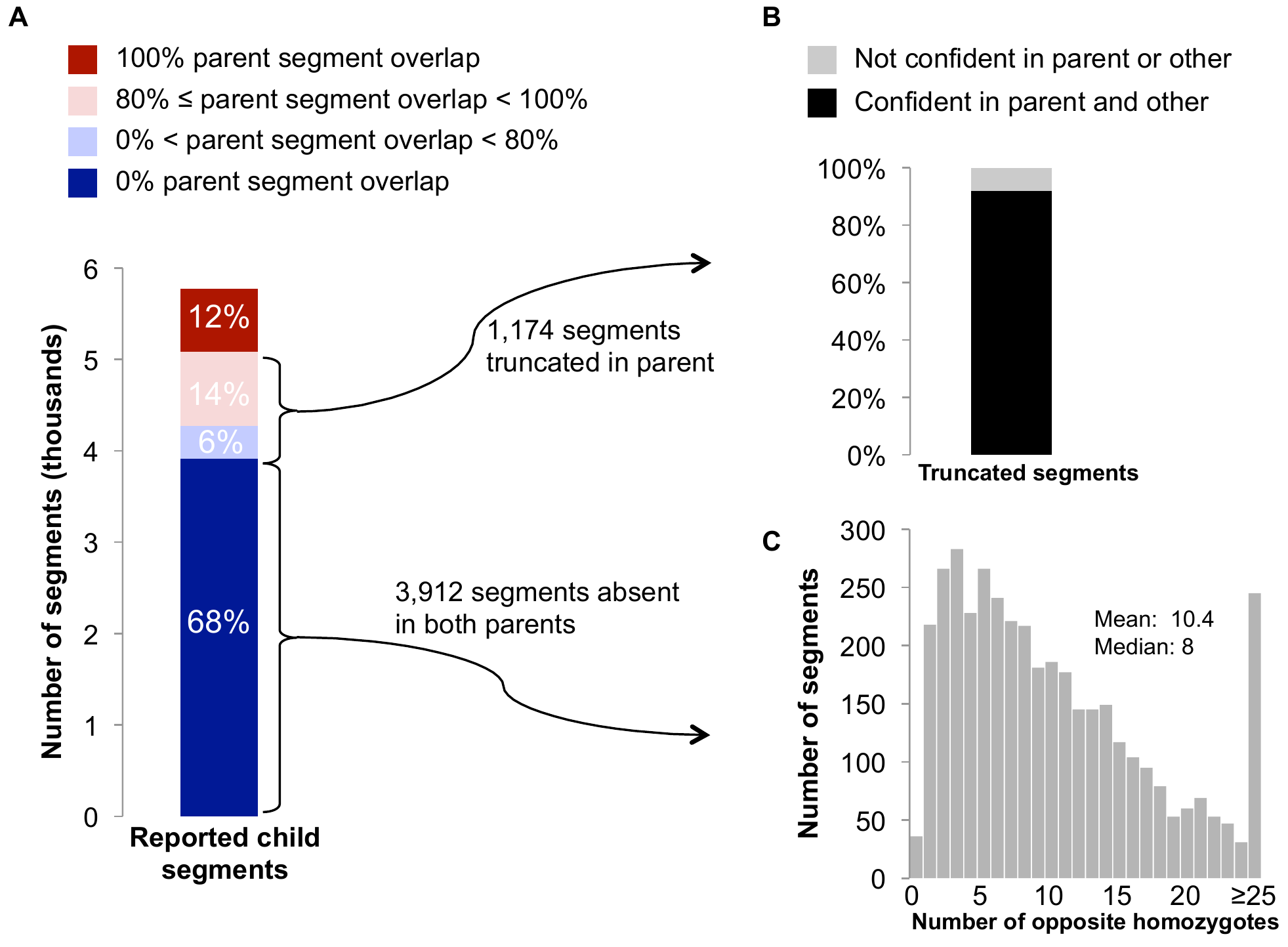}
\end{center}
\caption{
\textbf{Analysis of child-other segments in parents in the 1000 Genomes cohort.}
This figure is analogous to \figref{fig:falseSegs} but performed on
the 1000 Genomes cohort.
\textbf{A.} The majority of child-other segments are not detected in either parent.
\textbf{B.} Truncation points for parent-other segments are nearly always
confidently-genotyped opposite homozygote sites, consistent with false positive 
IBD in the child. The opposite homozygote site causing truncation of the 
parent-other segment was examined in all 1,174 segments with partial parent
overlap.
\textbf{C.} Child-other segments with no corresponding parent-other segments contain many 
parent-other opposite homozygotes in the region, also consistent with false 
positive IBD in the child. For each of these child-other segments, 
the number of opposite homozygote sites present between the parent and the other 
individual at that segment location is calculated separately for each parent, 
and the smaller is chosen as the number of opposite homozygotes in the region.
}
\label{fig:falseSegs1000G}
\end{figure}

\begin{figure}[!ht]
\begin{center}
\includegraphics[width=4in]{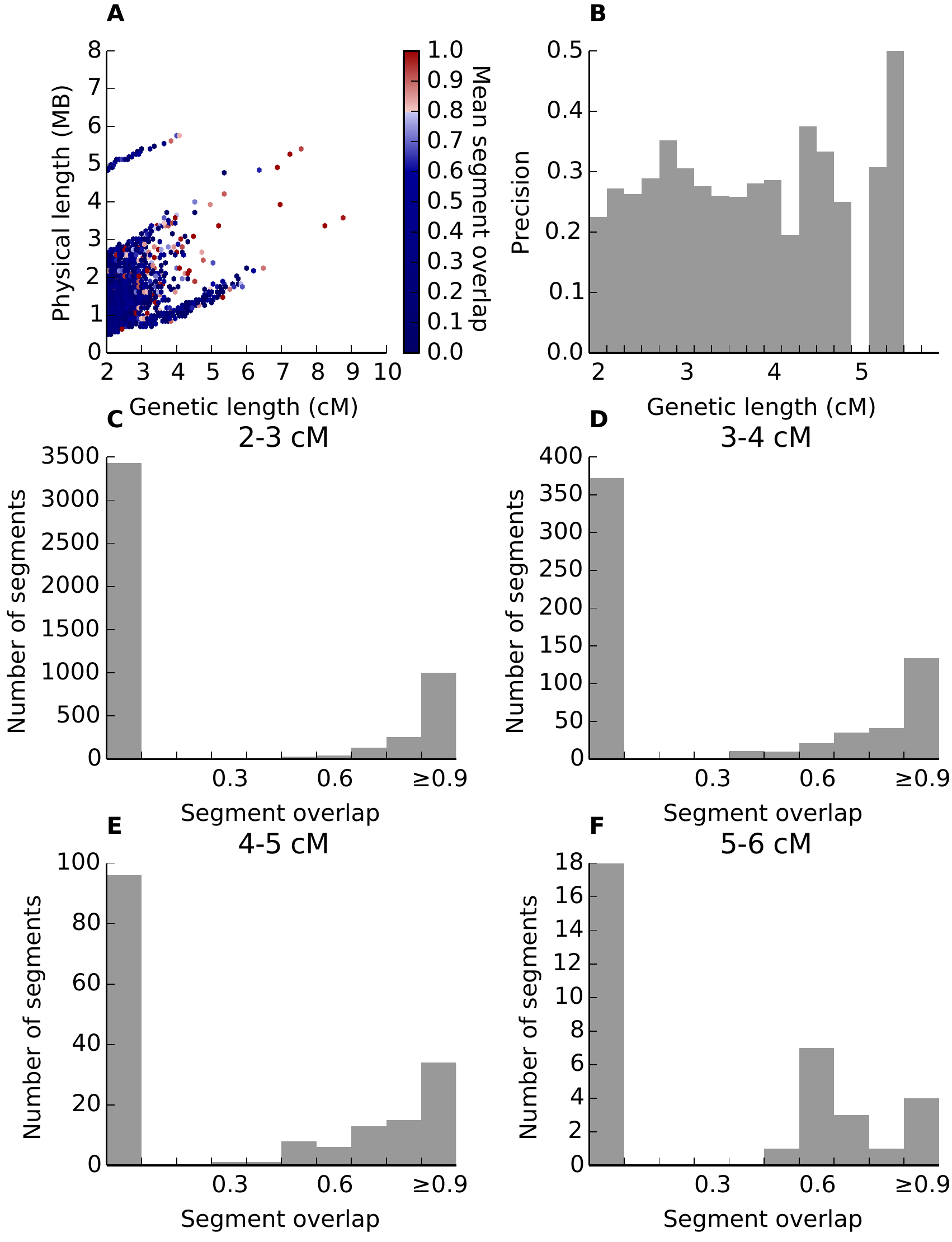}
\end{center}
\caption{
\textbf{Accuracy of child-other IBD segments reported by GERMLINE in the 1000 Genomes cohort.}
This figure is analogous to \figref{fig:originalPerformance} but performed on
the 1000 Genomes cohort.
\textbf{A.} Heat map of the mean fraction of reported child-other IBD segments 
contained in a corresponding parent-other segment, binned by two measures of
segment length as described in \figref{fig:originalPerformance}\textbf{A}.
\textbf{B.} The fraction of child-other segments that are true IBD as a function of
segment length. True IBD segments are defined as having at least 80\% of their
sites encompassed by a parent-other segment as in \figref{fig:originalPerformance}\textbf{B}.
\textbf{C--F.} Histograms of child-other segment counts binned by segment
overlap for segments of 2--3 cM (\textbf{C}), 3--4 cM (\textbf{D}), 4--5 cM
(\textbf{E}), and 5--6 cM (\textbf{F}).
Note the scale changes on the y-axes:
though the fraction of true segments of length $<3$ cM is smallest, this range
contains over 5-fold more true segments than all other length ranges 
combined.
}
\label{fig:originalPerformance1000G}
\end{figure}

\begin{figure}[!ht]
\begin{center}
\includegraphics[width=3in]{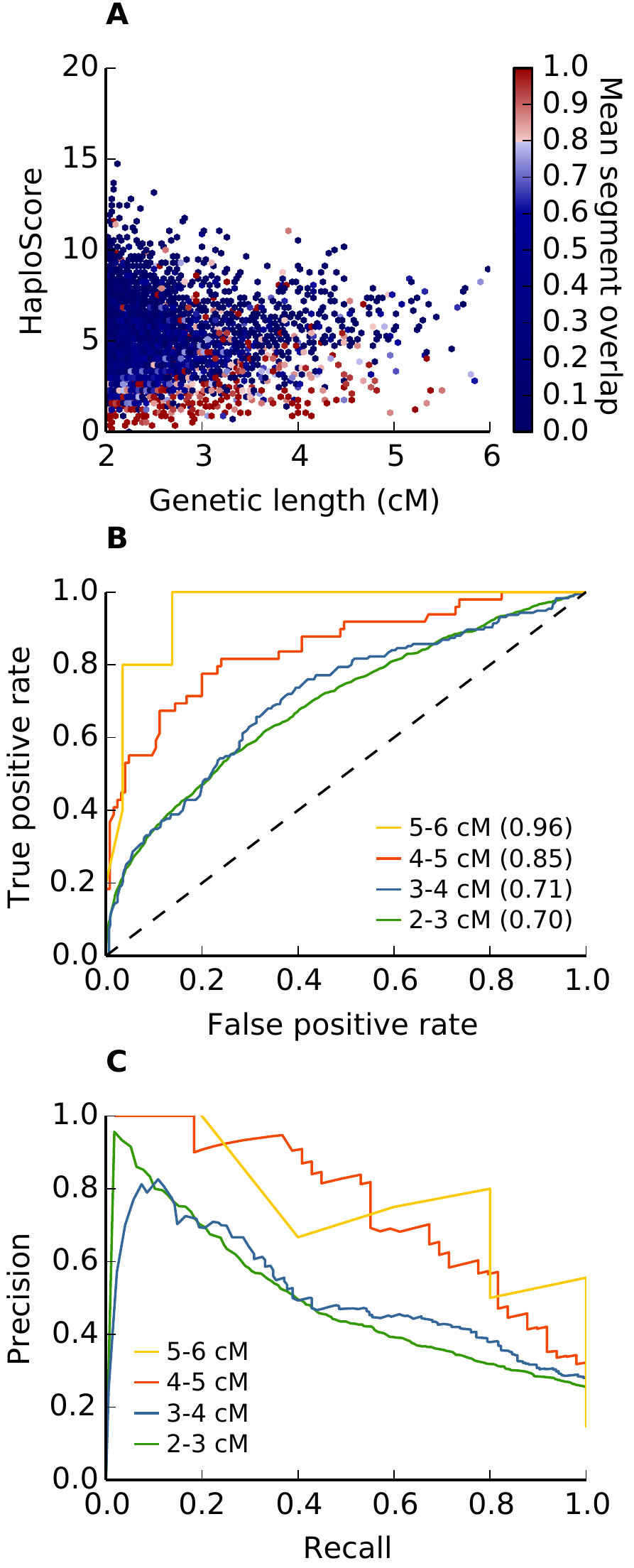}
\end{center}
\caption{
\textbf{Improving detection of true IBD segments using HaploScore in the 1000 Genomes cohort.}
This figure is analogous to \figref{fig:haploscorePerformance} but performed on
the 1000 Genomes cohort.
\textbf{A.} Heat map of the mean fraction of reported IBD segments found in 
parents, binned by segment genetic length and HaploScore. Calculations are
performed as in~\figref{fig:originalPerformance}\textbf{A}.
\textbf{B.} Receiver operating characteristic for reported IBD segments of
various lengths, discriminating by HaploScore. True IBD is defined as 
in~\figref{fig:originalPerformance}\textbf{B}. The dashed black line indicates 
the no-discrimination line. The area under each curve is parenthesized in its 
legend entry.
\textbf{C.} Precision-recall plot for child-other segments binned by segment
length.
}
\label{fig:haploscorePerformance1000G}
\end{figure}

\begin{figure}[!ht]
\begin{center}
\includegraphics[width=4in]{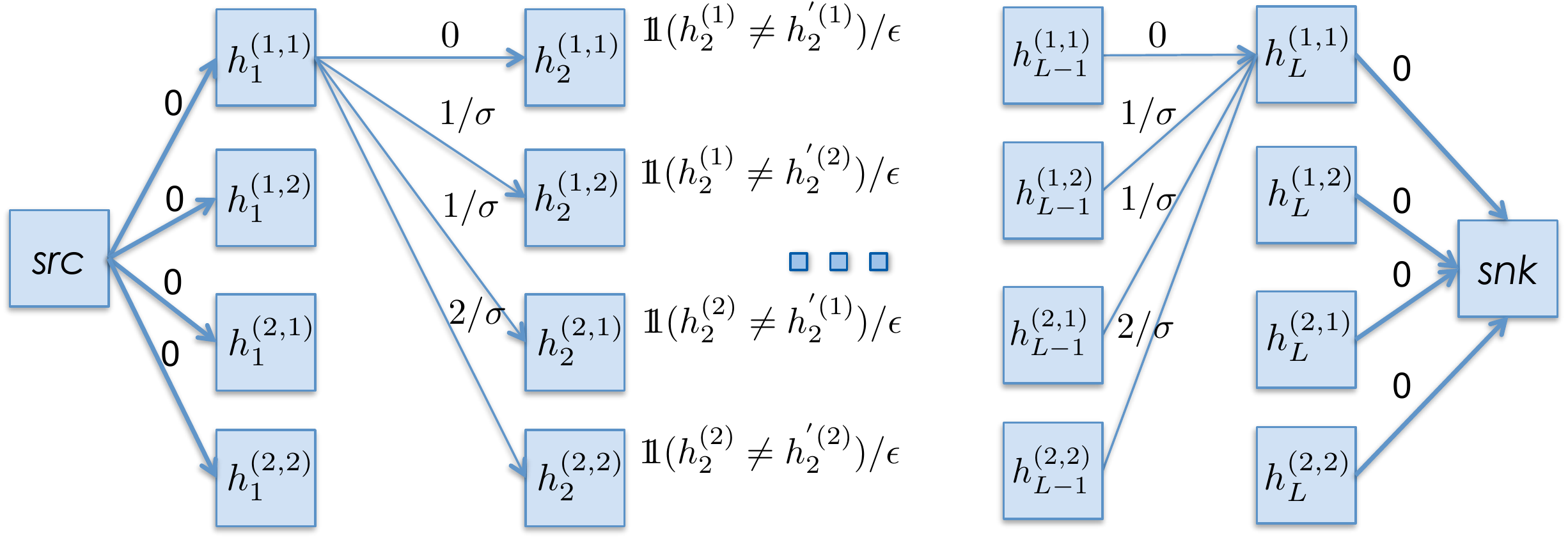}
\end{center}
\caption{
\textbf{Graph illustrating the HaploScore computation.}
The graph has one source, one sink and one level per genotyped site in the IBD
segment. At each level $l$, the graph contains four nodes, indicating the 
haplotype configuration at site $l$. Each node has weight 0 if the two 
corresponding alleles are the same, or $1/\epsilon$ if they are different. Each 
node in level $l$ has four outgoing directed edges, one to each of the four 
nodes in level $l+1$. The edge weights are $0$, $1/\sigma$, or $2/\sigma$, 
depending on whether 0, 1 or 2 switch errors are necessary to explain the 
transition. For clarity, some edges are omitted in this figure. The source node
$src$ has four outgoing directed edges with weight 0, one to each of the four 
nodes in level 1. Each node in level $L$ has one outgoing directed edge to the
sink node $snk$ with weight 0.
}
\label{fig:haploGraph}
\end{figure}
\clearpage

\clearpage
\begin{samepage}
\section*{Supplementary Tables}
\begin{table}[!ht]
\caption{
\textbf{Characteristics of the individuals in the 1000 Genomes cohort.}
}
\begin{tabular}{cccl}
\textbf{Population$^a$} & \textbf{Total trios} & \textbf{Total individuals} & \textbf{Reported relationships} \\
\hline
IBS & 50 & 150 & All trios \\
CEU & ~2 & 104 & Two trios, 98 unrelated$^b$ \\
GBR & ~0 & 101 & One mother-daughter pair, one unknown \\
 &  &          & second order relationship, 97 unrelated \\
TSI & ~0 & 100 & One sibling pair, 98 unrelated \\
FIN & ~0 & 100 & All unrelated  \\
\end{tabular}
\begin{flushleft}
$^a$ IBS, Iberian populations from Spain;
CEU, Utah residents with ancestry from northern and western Europe;
GBR, British from England and Scotland;
TSI, Toscans from Italy;
FIN, Finnish from Finland. \\
$^b$ Reportedly unrelated NA06989 and NA12155 share 35 cM on chromosome 21.
\end{flushleft}
\label{tab:cohort1000G}
\end{table}
\end{samepage}
\clearpage

\begin{table}[!ht]
\caption{
\textbf{Haplotype and diplotype window matches in child-other segments of 1000 Genomes data.}
\textbf{A.} Counts of window types in windows contained within a corresponding
parent-other segment.
\textbf{B.} Counts of window types in windows that are not contained within a
corresponding parent-other segment.
}
{\large \textbf{A}} \\
\begin{tabular}{lrrrr}
                  & \textbf{Child Diplo} & \textbf{Child Haplo} & \textbf{Child Both} & \textit{\textbf{Total}} \\
\hline                  
\textbf{Par None}      &  ~~~~0          &       ~~~0           &         ~~~~0       & \textit{~~~~~0}  \\
\textbf{Par Diplo}     &  3,257          &       ~~53           &          ~798       & \textit{~4,108}  \\
\textbf{Par Haplo}     &  ~~~50          &        167           &          ~170       & \textit{~~~387}  \\
\textbf{Par Both}      &  ~~817          &        169           &         7,702       & \textit{~8,688}  \\
\hline
\textit{\textbf{Total}}  & \textit{4,124} & \textit{389}        & \textit{8,670}      & \textit{13,183}  \\
\end{tabular} \\
{\large \textbf{B}} \\
\begin{tabular}{lrrrr}
                  & \textbf{Child Diplo} & \textbf{Child Haplo} & \textbf{Child Both} & \textit{\textbf{Total}} \\
\hline
\textbf{Par None}   &  ~7,955            &          ~~424       &         ~4,397      &   \textit{12,776}  \\
\textbf{Par Diplo}  &  ~6,037            &          ~~102       &         ~1,914      &   \textit{~8,053}  \\
\textbf{Par Haplo}  &  ~~~~90            &          ~~281       &         ~~~387      &   \textit{~~~758}  \\
\textbf{Par Both}   &  ~1,267            &          ~~278       &         ~8,227      &   \textit{~9,772}  \\
\hline
\textit{\textbf{Total}} & \textit{15,349} & \textit{1,085}      & \textit{14,925}     &   \textit{31,359}  \\
\end{tabular}
\begin{flushleft}
Par, parent; Diplo, diplotype match only; Haplo, haplotype match only.
\end{flushleft}
\label{tab:windowAnalysis1000G}
\end{table}
\clearpage

\begin{table}[!ht]
\caption{
\textbf{Genotype probabilities for a pair of individuals for different IBD states.}
}
\begin{tabular}{cccc}
\textbf{$(G_1, G_2)$} &  \textbf{IBD0}   & \textbf{IBD1}         & \textbf{IBD2} \\
\hline
AA, BB             &  $2 p^2 q^2$  &   0                &    0       \\
AA, AA             &  $p^4$        &   $p^3$            &    $p^2$   \\
AA, AB             &  $4 p^3 q$    &   $2 p^2 q$        &    0       \\
AB, AB             &  $4 p^2 q^2$  &   $p^2 q + p q^2$  &    $2pq$   \\
\end{tabular}
\begin{flushleft}
$p$ represents the allele frequency of allele A and $q$ $(=1-p)$ represents the 
allele frequency of allele B.
\end{flushleft}
\label{tab:ibdgenoprobs}
\end{table}
\clearpage

\begin{table}[!ht]
\caption{\textbf{Observed genotype probabilities with genotyping errors.}}
\begin{tabular}{cccc}
                   &  \textbf{$G_{true}=AA$}       &  \textbf{$G_{true}=AB$}            &  \textbf{$G_{true}=BB$}       \\
\hline
\textbf{$G_{obs}=AA$} &  $(1-\epsilon)^2$          &  $(1-\epsilon) \epsilon$        &  $\epsilon^2$              \\
\textbf{$G_{obs}=AB$} &  $2(1-\epsilon) \epsilon$  &  $(1-\epsilon)^2 + \epsilon^2$  &  $2(1-\epsilon) \epsilon$  \\
\textbf{$G_{obs}=BB$} &  $\epsilon^2$              &  $(1-\epsilon) \epsilon$        &  $(1-\epsilon)^2$          \\
\end{tabular}
\label{tab:observedgenoprobs}
\end{table}

\end{document}